\pdfoutput=1

\documentclass[%
reprint,
showpacs,
 amsmath,amssymb,
 aps,
 pre,
]{revtex4-1}
\usepackage[caption=false]{subfig}

\usepackage{graphicx}% Include figure files
\usepackage{dcolumn}% Align table columns on decimal point
\usepackage{bm}% bold math

\usepackage{amsmath}
\usepackage{amssymb}

\usepackage[linesnumbered,ruled]{algorithm2e}
\usepackage{hyperref}
\usepackage{xcolor}
\makeatletter
\Hy@AtBeginDocument{%
  \def\@pdfborder{0 0 1}% Overrides border definition set with colorlinks=true
  \def\@pdfborderstyle{/S/U/W 1}% Overrides border style set with colorlinks=true
}
\makeatother
\hypersetup{%
  colorlinks=true,% hyperlinks will be coloured
  linkcolor=blue,% hyperlink text will be xxx
  linkbordercolor=red,% hyperlink border will be xxx
}

\usepackage{lineno}
\usepackage{setspace}
\usepackage{microtype}
\DisableLigatures[f]{encoding = *, family = * }

\usepackage{amsfonts,url,float,color}

\usepackage{overpic}
\usepackage{footmisc}

\makeatletter
\renewcommand{\@biblabel}[1]{\quad#1.}
\makeatother

\date{}

\begin{document}
% \modulolinenumbers[3]
% \linenumbers 

\title{Motion Blur Filtering: {\emph{A Statistical Approach for Extracting  Confinement Forces and Diffusivity from a Single Blurred Trajectory}}}% Force 

\author{Christopher P. Calderon}
 \email{Chris.Calderon@UrsaAnalytics.com}
\affiliation{Ursa Analytics, Inc.}
\date{\today}% It is always \today, today,

\begin{abstract}

Single Particle Tracking (SPT) can aid in understanding a variety of complex spatio-temporal processes.  However, quantifying diffusivity and  confinement forces  from individual live cell trajectories is complicated by inter- \& intra-trajectory kinetic heterogeneity, thermal fluctuations, and (experimentally resolvable) statistical temporal dependence inherent to the underlying molecule's time correlated  confined dynamics experienced in the cell. The problem is further complicated   by experimental artifacts such  as localization uncertainty  and motion blur. The latter is caused by the tagged molecule emitting photons at different spatial positions during the exposure time of a single frame. The aforementioned experimental artifacts induce spurious time correlations in measured SPT time series that obscure the information of interest (e.g., confinement forces and diffusivity). 
We  develop a new maximum likelihood estimation (MLE) technique that  decouples the above noise sources and systematically treats temporal correlation  via time series methods.  This ultimately permits a reliable algorithm for  extracting diffusivity and effective forces in confined or unconfined environments.   
We illustrate how our approach avoids   complications inherent to mean square displacement (MSD) or autocorrelation techniques. 
 Our algorithm modifies the established Kalman filter (which does not handle motion blur artifacts) to provide a likelihood based time series estimation procedure.  The result extends Berglund's motion blur model [PRE, \textbf{82} (2010)]
 to handle confined dynamics.   The approach can also systematically utilize (possibly time dependent) localization uncertainty estimates afforded by image analysis if available.  
 To our knowledge, this is the first technique explicitly treating confinement and motion blur within a time domain MLE framework that uses an exact likelihood (time domain methods facilitate analyzing non-stationary signals).  
 Our new estimator is demonstrated to be consistent over a wide range of exposure times ($5 - 100 \ ms$), diffusion coefficients $(1\times10^{-3}$  -  1 $\mu m^2/s)$, and confinement widths  ($100 nm - 2\mu m$).  We demonstrate that neglecting motion blur or confinement can substantially bias estimation of kinetic parameters 
of interest to researchers. 
The technique also permits one to check statistical model assumptions  against measured individual trajectories without ``ground truth''.
 The ability to reliably and consistently extract motion parameters  in trajectories exhibiting confined and/or non-stationary dynamics , without exposure time artifacts corrupting estimates,   is expected to aid in directly comparing trajectories obtained from different experiments or imaging modalities. 
A Python implementation is provided; open-source code will be maintained on \href{http://www.github.com/calderoc/MotionBlurFilter}{GitHub}
\footnote{This version differs in minor typographical details from the published version, DOI:10.1103/PhysRevE.93.053303, available on-line at:\url{http://journals.aps.org/pre/abstract/10.1103/PhysRevE.93.053303}.  Code will be maintained at \url{http://github.com/calderoc/MotionBlurFilter}.}. 

\end{abstract}
 
\maketitle 
\renewcommand*{\thefootnote}{\arabic{footnote}}
\setcounter{footnote}{0}

 \section{Introduction}

The number of techniques available to accurately probe molecules in their native, crowded, and time changing live cell environment  has increased dramatically in recent years
\cite{Arhel2006,Brandenburg2007,Lange2008,Manley2008,Joo2008b,Biteen2011,Danuser2011,Ram2012,Meijering2012,DeWit2012,Pawar2014,Polo2011,Rohatgi2007,Hajjoul2013,Ye2013,Gahlmann2013,Levi2005,Pavani2009,Wells2010,Thompson2010,Ashley2012,Backlund2014,Vasquez2014,Welsher2014,Chen2014,Li2015,Wu2015}.  Research aimed at more efficiently extracting kinetic information from live cell single particle tracking (SPT) experiments has also experienced rapid growth \cite{Montiel2006,Berglund2010,Weber2012,Persson2013,Calderon2013,Calderon2013b,Hoze2014,Calderon2014,VandeMeent2014,Vestergaard2014,Backlund2015,Beheiry2015,Hines2015,CalderonBloom2015,Monnier2015}.  However, analysis methods have substantially lagged behind microscopy developments. An important and ubiquitous problem in cell biology \cite{Hofling2013,Metzler2014} that has not received substantial statistical attention is associated with how to address various technical challenges inherent to analyzing confined motion \cite{Kusumi1993,Levi2005,Destainville2006} in a collection of experimental trajectories exhibiting heterogeneous and/or non-stationary (i.e., transient kinetic phenomena)  responses \cite{Saxton1997,Chen2013a,KrapfCCP2013,Calderon2013b,Metzler2014,CalderonBloom2015}.  
High-resolution multicolor image stacks can provide hints of molecular interactions when analyzed via a spatial 
co-localization  analysis \cite{Wu2015}, however reliably distinguishing between transient molecular binding events (hence changing the underlying molecular diffusivity of the biomolecule) 
vs. 
 coincidental co-localization 
 can be aided by new quantitative time series methods.
 Measurement apparatus noise further complicate the problem, since  in optical microscopy, position measurements are  not ``instantaneously" observed.  Empirically measured data is a time averaged quantity \cite{Savin2005,Destainville2006,Berglund2010}.  For example, in fluorescence based optical microscopy  \cite{Danuser2011,Gahlmann2013,Liu2015},
 position is
inferred from the observed Point Spread Function (PSF) \cite{Thompson2002,Ober2004,Gahlmann2013,Liu2015} obtained by collecting a finite number of photons emitted as the tagged molecule moves throughout the cell \cite{Savin2005,Destainville2006,Berglund2010,Gahlmann2013,Backlund2015,Ashley2015}.  The noise due solely to photon emissions from multiple positions in a single image is what we refer to  generically as ``motion blur''. 
The statistical correlation between thermal fluctuations, confinement forces, and motion blur introduces new time series challenges not addressed by current SPT data analysis routines.  
Refs. \cite{Berglund2010,Vestergaard2014} address how to handle  motion blur issues under the assumption of simplified diffusion models, but fail to address technical complications associated with confined motion  (a common occurrence in live cell SPT studies).  The classic Kalman Filter (KF) algorithm can deal with simple confined motion models, but
fundamental assumptions behind the KF are violated when motion blur is present \cite{hamilton,anderson1979}.

We introduce a new likelihood based estimation scheme 
that explicitly models  (i) the spatio-temporal statistical correlation inherent to molecular position measurements undergoing confined diffusion \cite{Destainville2006,Calderon2013}; and (ii) the correlation between various sources of measurement noise  and the underlying particle position time series
in camera based measurements (the approach treats  ``dynamic"  and ``static" measurements errors commonly encountered in SPT \cite{Backlund2015}).
 The technique is capable of reliably estimating local molecular diffusivity, $D$, and instantaneous  velocity \& forces from a single noisily measured position vs. time SPT trajectory.  \emph{Reliably estimating the aforementioned kinetic quantities  
  from  SPT trajectory measurements requires one to address statistical correlation induced by confinement, localization, and motion blur \cite{Destainville2006}.}  Obtaining reliable estimates of instantaneous forces from a single trajectory and respecting heterogeneity commonly encountered in SPT data \cite{Saxton1997,Chen2013a,KrapfCCP2013,Calderon2013b,Metzler2014,CalderonBloom2015} requires an accurate estimate of $D$ that is free of measurement apparatus artifacts  
  (our approaches provides this information).
 Our approach also enables researchers to use a single measured trajectory produced by an individual molecule (where localization quality may vary over time) 
 to systematically decouple ``dynamic measurement errors'' from effective ``static measurement errors'' \cite{Savin2005,Destainville2006,Backlund2015}. This decomposition is described in detail later since it is a key technical aspect of our new algorithm.
 
As we demonstrate, failing to properly account for motion blur and/or confinement can substantially affect quantities required to estimate forces from SPT trajectories. The Motion Blur Filter (MBF) algorithm addresses the technical concerns
through a reformulation of the KF \cite{hamilton,anderson1979}.
To our knowledge, this is the first time domain likelihood based approach that explicitly accounts for  confinement and motion blur. 
Likelihood-based approaches accounting for the natural time ordering of measurements \cite{SPAfric,SPAdsDNA,Calderon2013b} are advantageous since transient (but experimentally resolvable) changes in molecular forces
cannot be readily detected by legacy approaches such as mean square displacement (MSD) \cite{Destainville2006} or autocorrelation  approaches \cite{Savin2005,Backlund2015} where implicit assumptions about stationary statistics are made.
The MBF algorithm utilizes closed-form analytical expressions from stochastic process theory and avoids \emph{ad hoc} statistical approximations to the likelihood function.  The ability of our technique to consistently estimate diffusion coefficients ranging from $1\times10^{-3}$  to 1 $\mu m^2/s$ in the presence of molecular confinement sampled using 
observations spaced ``finely'' (5 $ms$) to ``coarsely'' (100 $ms$) in time is demonstrated via simulations.  

Our approach does assume that some approximate localization technique \cite{Thompson2002,Ober2004,Abraham2010,Gahlmann2013,Liu2015}, e.g. centroid or PSF shape based,  
can 
extract an unbiased  estimate of the time averaged position (where averaging occurs over the exposure time of a single image)  of the molecule of interest;  note that the ``quality "of the localization is permitted to vary  over time within our framework. 
A data-driven technique, processing``blurred'' data,  capable of decoupling various noise sources over a wide range of exposure times  
is expected to aid in identifying physically relevant motion experienced \emph{in vivo} in a variety of dynamic processes.  
The MBF provides an algorithm enabling a more direct and reliable comparison of  parameters obtained from data with different temporal and spatial resolution since biases affecting other methods at different length and time scales are not experienced by the MBF.  This is expected to result in a more comprehensive picture of dynamics occurring in the cell.

Note that when biomolecules are sampled with single-molecule precision \emph{in vivo}, they often experience transitions from ``standard diffusion'' to ``anomalous diffusion"  as the timescale of observation increases \cite{Hofling2013}. 
A longer term aim of this work is to provide a statistically robust method capable of accurately quantifying the motion parameters associated with single-molecule data 
before events leading to ``anomalous diffusion" phenomena manifest and can be statistically detected within individual trajectories.  Our motivation is to extract ``finer scale'' molecular kinetic information from the high temporal and spatial resolution measurements afforded by contemporary optical microscopy with the hope of providing a tool which can accelerate detection of new dynamic phenomena from these measurements \cite{Arhel2006,Brandenburg2007,Lange2008,Manley2008,Biteen2011,Danuser2011,Ram2012,Meijering2012,DeWit2012,Pawar2014,Polo2011,Rohatgi2007,Hajjoul2013,Ye2013,Gahlmann2013,Levi2005,Pavani2009,Wells2010,Thompson2010,Ashley2012,Backlund2014,Vasquez2014,Welsher2014,Chen2014,Li2015,Wu2015}.

This article is organized as follows:  Section \ref{sec:qual} introduces the models, theoretical background, and the new MBF algorithm.   
%In subsection \ref{sec:qualitativeExamples},
Towards the end of this section (in subsection \ref{sec:qualitativeExamples}), 
we  walk the reader through two examples illustrating how to use the open-source Python code and IPython Notebooks  provided along with this article (these examples  reproduce Figs. \ref{fig:nonstatVelocityEG} and \ref{fig:RvaryingEG}).
 Section \ref{sec:results} presents  additional results on large scale simulations and Section \ref{sec:conclusion} concludes.  An Appendix provides additional mathematical details.  The Supp. Mat. \footnote{See Supplemental Material at [URL
  will be inserted by publisher] for additional results and  algorithmic  details.} 
  provides additional results and  algorithmic  details.

\section{Model and Methods}
\label{sec:qual}

  This subsection is organized as follows:  the assumed continuous time dynamical model is introduced in Sec. \ref{sec:notation}.   Prior to discussing the finer technical time series and filtering details, we expand on the assumptions made about the measurement noise sources in Sec. \ref{sec:measErrorQual}, since these assumptions are key to understanding the MBF.   After a detailed description of the measurement noise, we briefly discuss physical interpretations of the model in Sec. \ref{sec:physicalSDE} (additional details can be found in Refs. \cite{Calderon2013b} and \cite{CalderonBloom2015}).
  This  discussion is followed by a review of the classic KF in Sec. \ref{sec:basicOfKF}.  After providing the required background, 
  in Sec. \ref{sec:methods} we introduce the MBF algorithm. 
  Subsequently, we provide two illustrative examples in Sec. \ref{sec:qualitativeExamples} illustrating advantages of the MBF and how one can leverage the approach on SPT data.  This section concludes by contrasting the MBF to existing approaches in SPT.

\subsection{Continuous Time Model with Discrete ``Blurred'' Measurements}
\label{sec:notation}

The MBF assumes the following Stochastic Differential Equation (SDE) and measurement model:

\begin{align}
\label{eq:SDE}
dr_t = & ({v}-{\kappa} r_t)dt + \sqrt{2 D}dB_t \\
\label{eq:contMeas}
\psi_{t_i} = & \frac{1}{t_E}\int_{t_{i}-t_E}^{t_i} r_s ds + \epsilon^{\mathrm{loc}}_{t_i}
\end{align}

  The true position of the tagged particle at time $t$ is denoted by $r_t$ \footnote{This subscript notation is standard in SDE modeling \cite{kp}.} and the discretely sampled position measured by the microscope is denoted by $\psi_{t_i}$.  The stochastic term driving the SDE above is a standard Brownian motion, denoted by $B_t$.  The effective ``static" \cite{Thompson2002,Ober2004,Abraham2010,Backlund2015,Ashley2015} localization measurement noise at ${t_i}$  measured by a camera with exposure time  ${t_E}$   is denoted 
 by $\epsilon^{\mathrm{loc}}_{t_i}$ (this noise is modeled as a mean zero Gaussian random variable independent of the particle position).  The time integral in  Eq. \ref{eq:contMeas} models ``motion blur"  introduced by the camera \cite{Berglund2010,Gahlmann2013}
 \footnote{The MBF was introduced in a form motivating by continuous illumination \cite{Berglund2010,Gahlmann2013}. The  basic approach MBF can  be modified to handle other ``shutter functions'' \cite{Berglund2010} enabling camera modes outside of continuous illumination, we focus on continuous uniform illumination since this mode is most commonly used in biophysics and cell biology 
  experiments \cite{Berglund2010,Gahlmann2013,Liu2015}.}.
 Additional assumptions and details regarding the measurement noise sources are discussed later in this section and in greater detail in  Sec. \ref{sec:measErrorQual}. 
The model above is specified by a parameter vector denoted by   
$\theta=(v,\kappa,D,{\sigma_{\mathrm{loc}}})$ (this parameter vector contains physically interpretable parameters described in Secs. 
\ref{sec:measErrorQual} - \ref{sec:physicalSDE}). 

The continuous time SDE in Eq. \ref{eq:SDE} is a linear model whose solution is known precisely \cite{risken,kp}.  Using $\kappa > 0 $ allows confinement  to be modeled (a common phenomenon in live cell studies). However, the spatial and temporal dependence introduced when $\kappa \ne 0 $ and motion blur corrupts observations poses several new time series challenges not  addressed previously  \cite{Berglund2010,Calderon2013b}.  
The MBF algorithm processes  discrete measurements samples taken from the continuous time SDE.
   The linear and Gaussian nature of Eq. \ref{eq:SDE}  enables an exact discretization of Eq. \ref{eq:SDE}, as well as a discrete representation of the integral appearing in Eq. \ref{eq:contMeas}  that does not introduce any statistical errors into the filtering framework (these features are crucial to the MBF).   The aforementioned discretization of Eqs. \ref{eq:SDE}-\ref{eq:contMeas} is given by:

\begin{align}
\label{eq:discreteSDE}
r_{t_{i+1}} = \ & A + F r_{t_{i}} + \eta_{t_i} \\
\label{eq:discreteMeas}
\psi_{t_i} = \ & H_A + H_Fr_{t_{i-1}} + \epsilon^{\mathrm{loc}}_{t_i} + \epsilon^{\mathrm{mblur}}_{t_i} \\
\label{eq:discreteSDEnoiseDist}
\epsilon^{\mathrm{loc}}_{t_i} + & \epsilon^{\mathrm{mblur}}_{t_i} \sim  \mathcal{N}(0,R_i) \\
\label{eq:discreteSDEMeasQ}
\eta_{t_i} \sim  & \mathcal{N}(0,Q) \\
\label{eq:discreteSDEMeasQblur}
 \epsilon^{\mathrm{mblur}}_{t_i}  \sim & \mathcal{N}(0,Q^{\mathrm{mblur}}) \\
\label{eq:discreteSDEcrossCorr}
  C = \ & \mathrm{cov}(\eta_{t_{i}},\epsilon^{\mathrm{mblur}}_{t_{i+1}}) \\
\mathrm{cov}(\eta_{t_{i}},\epsilon^{\mathrm{mblur}}_{t_{j}}) = & 0 \ \mathrm{if } j \ne i+1 \\
\label{eq:discreteSDEnoCorr}
 \mathrm{cov}(\eta_{t_{i}},\epsilon^{\mathrm{loc}}_{t_{j}}),  & 
 \mathrm{cov}(\epsilon^{\mathrm{loc}}_{t_{i}},\epsilon^{\mathrm{mblur}}_{t_{j}}) =  0 \ \forall i,j 
  \end{align}

\noindent where 
 $\mathcal{N}(\mu,\sigma^2)$ denotes a Gaussian random variable with mean $\mu$ and variance $\sigma^2$.
The quantities $(A,C,F,H_A,H_F,Q,Q^{\mathrm{mblur}},R_i)$ listed in the discrete equations above  can be 
derived in closed-form  given $\delta_i := t_{i} - t_{i-1}$, $t_E$, and the continuous time parameters  $\theta$ (expressions not explicitly defined in this section are derived and provided in the Appendix).  The term  $\eta_{t_i}$ represents a Gaussian mean zero  ``process noise" \cite{hamilton,anderson1979} with variance $Q$; the term  ``process noise'' is used in control theory to describe a stochastic noise source that affects the true underlying state of the system (the ``state" is molecular position in the application considered here).
The 
term $\epsilon^{\mathrm{mblur}}_{t_i}$ 
represents the difference between $r_{t_i}$ and the conditional expectation 
of $\frac{1}{t_E}\int_{t_{i}-t_E}^{t_i} r_s ds$ (conditioned on $r_{t_{i-1}}$).

  An important technical aspect of the discretized model above is  the fact that 
the conditional expectation $\mathbb{E}[\frac{1}{t_E}\int_{t_{i}-t_E}^{t_i} r_s ds|r_{t_{i-1}}]$ is a Gaussian random variable with
mean $H_A + H_Fr_{t_{i-1}}$ and variance $Q^{\mathrm{mblur}}$. The parameters $H_A$ and $H_F $ are used to compute the expected value of the measurement at the next time instant (since there is motion blur, the mean of the measurement does not coincide with the value of the underlying position at the same time point).  The parameters $A$ and $F$ play a similar role, i.e., 
$\mathbb{E}[r_{t_i}|r_{t_{i-1}}]$  = $A + Fr_{t_{i-1}}$ ($A$ and $F$  type parameters are commonly encountered in KF applications when a continuous time model is inferred from discretely observed measurements \cite{Calderon2013}).

Equation \ref{eq:discreteSDEcrossCorr} emphasizes that $\epsilon^{\mathrm{mblur}}_{t_{i+1}}$ \emph{is statistically correlated} with the $\eta_{t_{i}}$ (all other noise terms above are statistically independent).
The net measurement noise variance obtained by combining the static and dynamic error is 
$R_i:= Q^{\mathrm{mblur}} + (\sigma_{\mathrm{loc}}^{{\mathrm{Input}}}(t_i) + \sigma_{\mathrm{loc}})^2$ (the term in parenthesis permits time dependent effective static measurement noise; these terms are described in detail in the next subsection \& Illustrative Example 2 in Sec. \ref{sec:qualitativeExamples}).
 Figure \ref{fig:twonoisetypes} illustrates the motion blur measurement noise  and effective static measurement noise sources graphically (these terms are further described in Sec. \ref{sec:measErrorQual}).

 Under the assumed parametric model (see Eqs. \ref{eq:SDE} and \ref{eq:contMeas}), the variance of $\epsilon^{\mathrm{mblur}}_{t_i}$
and its covariance with $r_{t_i}$ (covariance quantified by $C$ in Eq. \ref{eq:discreteSDEcrossCorr}) can be obtained in closed-form in terms of the model parameters (derivation provided in Appendix).    The classic Kalman Filter (KF) can account for confinement forces and non-stationary statistics \cite{Calderon2013b}, but the classic KF is not able to explicitly handle motion blur due to the correlation between $\epsilon^{\mathrm{mblur}}_{t_i}$ and $\eta_{t_{i-1}}$.
 \emph{Both the practical utility and main technical contribution of the MBF are associated with how the MBF handles Eq. \ref{eq:contMeas} and the correlation in Eq. \ref{eq:discreteSDEcrossCorr}}, hence a detailed qualitative description of the physical phenomena underlying the assumptions behind the ``measurement error''  terms  is presented before proceeding.

\subsection{Qualitative Description of Static and Dynamic Measurement Errors}
\label{sec:measErrorQual}

\begin{figure}[htp]
\center
\centering
\begin{minipage}[b]{1.\linewidth}
\def\pw{1}
% \begin{overpic}[width=\pw\textwidth]{./Fig1New.pdf}
\begin{overpic}[width=\pw\textwidth]{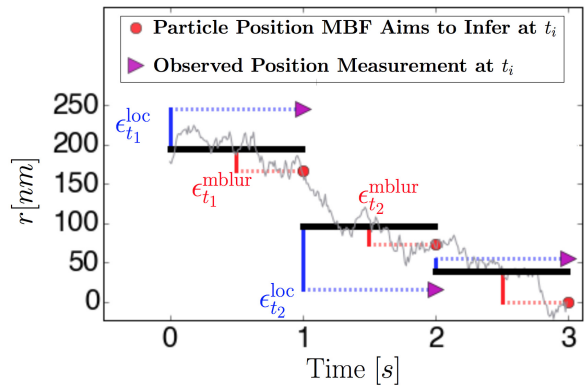}
\end{overpic}
\end{minipage}
\caption{\footnotesize \textbf{Illustration of ``Dynamic"  and Effective ``Static"  Measurement Errors.}  
The solid gray line represents the true (unobservable) position trajectory  of the molecule evolving in continuous time.  The trajectory measured without noise contains temporal autocorrelation due to confinement forces \cite{Savin2005,Destainville2006,Berglund2010,Backlund2015}. 
The thick black horizontal lines denote the time averaged position recorded during the camera's exposure time,  $t_E$.   The Motion Blur Filter (MBF) aims  to infer  the position at time $t_i$ (denoted by red circles) from a sequence of noisy measurements,  $\{\psi_{t_i}\}_{i=1}^T$ (denoted by purple triangles).  Each measurement, $\psi_{t_i}$,  is assumed to come from a Point Spread Function (PSF) generated by a single molecule emitting a finite number of photons at different positions during image acquisition 
(``image $i$ "is measured during the time interval $[t_{i}-t_E,t_i]$).  This induces what we refer to as ``motion blur'' measurement error  
and this noise
is denoted by $\epsilon^{\mathrm{mblur}}_{t_i}$. The idealized effective static 
localization error (induced by finite photon counts in PSF estimation) is denoted by $\epsilon^{\mathrm{loc}}_{t_i}$. See text for additional details on the assumptions behind these measurement noise sources.  
}
\label{fig:twonoisetypes}
\end{figure}

In SPT analysis of fluorescence optical microscopy data, the position measurements of a single molecule are typically obtained from through localization techniques \cite{Thompson2002,Ober2004,Abraham2010,Backlund2015,Ashley2015}.  Throughout the remainder of this article, the term ``PSF'' is used to represent the observed Point Spread Function. The inference of the PSF can be guided by optics principles, but we use the term ``PSF'' to represent a data-driven quantity extracted from an image measured by an optical light microscope.    If the molecule being imaged by a microscope is mobile, the observed PSF is contaminated by both ``static" and ``dynamic" measurements errors \cite{Savin2005,Destainville2006,Berglund2010,Backlund2015}.  The integral in Eq. \ref{eq:contMeas} models the ``dynamic" motion blur measurement error, $\epsilon^{\mathrm{mblur}}_{t_i}$,  due to the tagged molecule emitting photons  at  different  spatial positions in the time interval producing a single image, $[t_{i}-t_E,t_i]$. The motion blur measurement error is highly correlated with the underlying particle position of interest.   The effective ``static" measurement error is intended to quantify the error associated with inferring a PSF's shape from observed pixel intensities  \cite{Thompson2002,Ober2004,Abraham2010,Backlund2015,Ashley2015}.  That is, the effective static measurement error is an idealized quantity (estimated from observational data) which aims at quantifying the error due  to estimating a PSF from a finite number of measured photons.  
The dynamic or immobile nature of the emitters producing the data used to calibrate the PSF is not relevant to the data-driven effective static measurement noise used by the MBF.  
Note that the observed PSF may not  be radially symmetric even in the ``ideal infinite photon'' limit due to the dynamic errors introduced by molecular motion.  However, as stated in the introduction, the MBF does assume that the localization procedure provides an unbiased point estimate of the the time averaged position of the molecule of interest in the image measurement (the mean zero assumption of $\epsilon^{\mathrm{loc}}$ in Eq. \ref{eq:discreteMeas} reflects this assumption).  
In an ideal ``infinite photon collection" scenario, the variance of $\epsilon^{\mathrm{loc}}$ would be zero.  However, even in the infinite photon case, there would still be uncertainty in the molecular position if the molecule moves and the exposure time is non-zero.  Both measurement errors are illustrated in Fig. \ref{fig:twonoisetypes}. 

The main assumptions underlying the MBF are that (i) the dynamics are consistent with  Eqs. \ref{eq:SDE}  and (ii) the error introduced by finite photon count PSF estimation can be approximated by a Gaussian random variable whose statistics are determined by Eq. \ref{eq:contMeas}.  
 The MBF framework  recognizes that within empirically observed measurements, static and dynamic measurement noise sources are convolved in the raw data.   
The MBF approach uses a data-driven  model-based approach to ``decouple'' static and dynamic errors. 
\emph{Although the position measurement is blurred  due to molecular motion during image acquisition and finite camera exposure time \cite{Savin2005,Destainville2006,Berglund2010,Backlund2015}, the MBF aims at rigorously inferring the instantaneous position of the molecule at $t_i$ as well as the parameters characterizing its motion}.   The SDE model-based framework outlined  in Eqs. \ref{eq:discreteSDE} - \ref{eq:discreteSDEnoCorr} is key to achieving the noise decomposition described in this section.  
In most SPT applications, the  Gaussian assumption on the effective static measurement noise  cannot typically be statistically rejected if 10 or more photons underlie a PSF estimate (formal means for detecting Poisson artifacts could be considered \cite{arxivDec2013}). 
 In this article, goodness-of-fit tests are used to detect if  data is consistent with various assumptions underlying the assumed dynamical models (the tests used  have been shown capable of detecting artifacts of ``low photon count" measurement error in SPT data if they are  present in the data \cite{arxivDec2013}).

In the open-source software provided (discussed in Sec. \ref{sec:qualitativeExamples}) with this article, we allow estimates of time dependent localization noise statistics through the optional input $\sigma_{\mathrm{loc}}^{{\mathrm{Input}}}(t_i)$. The time dependent input \emph{estimate} of the localization uncertainty at time $t_i$ can be biased (e.g., it may contain artifacts of motion blur).  When the optional sequence $\sigma_{\mathrm{loc}}^{{\mathrm{Input}}}(t_i)$ is input, the MBF 
  uses the full sequence of measurements to estimate the parameter $\sigma_{\mathrm{loc}}$ which defines the net effective time dependent static measurement  noise variance:
  $(\sigma_{\mathrm{loc}}^{{\mathrm{Input}}}(t_i) + \sigma_{\mathrm{loc}})^2$ for each $i$.  It is stressed that the net \emph{effective} static measurement error is estimated from the data and not assumed known \emph{a priori} (the parameter $\sigma_{\mathrm{loc}}$ adjusts for the fact that the input localization estimates are likely corrupted by motion blur or other artifacts).  
  If $\sigma_{\mathrm{loc}}^{{\mathrm{Input}}}(t_i)$ is not provided as input, the estimated parameter $\sigma_{\mathrm{loc}}$ denotes the constant effective static measurement noise associated with the time series data.

\subsection{Physical Interpretation of Continuous Time SDE Parameters}
\label{sec:physicalSDE}
Recall that the SDE of interest (Eq. \ref{eq:SDE}) is characterized by the parameter vector  
$\theta=(v,\kappa,D,{\sigma_{\mathrm{loc}}})$ where  $D$ denotes the local effective diffusion coefficient; $\kappa$ and $v$ characterize the  instantaneous velocity  \cite{Calderon2013b,Calderon2014,CalderonBloom2015}.   The parameter $\sigma_{\mathrm{loc}}$ was described in detail in the previous section. In live cell studies, motion is often confined and confinement affects the temporal correlation statistics \cite{Destainville2006} even if motion blur and measurement artifacts are not present in the data.
The SDE model considered can use a single trajectory to compute ``the instantaneous force" from the estimated diffusion coefficient \cite{CalderonBloom2015}.    The ability to use a single trajectory to estimate motion parameters 
permits researchers to quantify heterogeneity and time changing forces at different points in the cell \cite{CalderonBloom2015}.   For example, the effective force at time $t$, denoted by $F(t)$, is approximated by $\frac{k_BT}{\hat{D}}(\hat{v}-\hat{\kappa} r_t)$ where hats 
denote the maximum likelihood estimate (MLE) extracted from the data.  In the previous force equation, we appealed to the classical Einstein relationship, i.e., $D = \frac{k_BT}{\gamma}$, where  $k_BT$ is Boltzmann's constant multiplied by the system temperature and $\gamma$ is the effective molecular friction \cite{risken,Calderon2013b}.  Our technique for estimating forces assumes that ``cage hopping'' or ``crowding'' events have  not occurred within the observed trajectory
\cite{Hofling2013,Chen2013a,Backlund2015,Holcman2015}.   Note that we are not restricting the term ``cage hopping'' to refer to kinetic phenomena in membrane diffusion  \cite{Kusumi1993,Destainville2006}.
In the presence of SPT trajectories spanning ``long times'', we acknowledge that crowding in the cell may result in cage hopping type phenomena. In Sec. \ref{sec:conclusion}, we  discuss techniques to preprocess trajectories by  segmenting the data into regimes where the MBF technique can be used to extract reliable force from position vs. time data using the model above as a building block.  If cage hopping type phenomena  occurs on time scales much faster than the temporal resolution afforded by the measurement device (e.g., many ``cage jumps'' can occur during the exposure time), the MBF can still be used to estimate effective SDE parameters explicitly accounting for the statistical effects of motion blur.

In the remainder of this subsection, we illustrate how the  SDE above nests other popular SPT models (i.e., we outline how directed and pure diffusion models are special cases of the SDE model in Eq. \ref{eq:SDE}). If both $v$ and $\kappa$ are set to zero, one obtains the motion blur model considered by Berglund \cite{Berglund2010} (handling the $v\ne 0,\kappa=0$, is a simple extension of Berglund's result as we show in the code supplied).
When $\kappa \ne 0$, one has an Ornstein-Uhlenbeck (OU) process \cite{risken}. If $\kappa > 0$ ($\kappa \ge 0$ is  assumed by the MBF), the OU  process can model confined diffusion.  A popular SPT confinement parameter, $L$ \cite{Kusumi1993}, is closely connected to the so-called ``corral radius''.  The original usage of the parameter $L$ in SPT corresponded to a box width (not a proper ``radius") of a hard wall potential \cite{Kusumi1993}. However, the ``coral radius'' term is often used in  SPT  to generically describe the square root of the asymptotic limit of an MSD curve of a confined particle \cite{Destainville2006,Park2010,Calderon2013}.   $L$ is related to the parameters of the model considered here through the equality $L = \sqrt{\frac{12D}{\kappa}}$ 
(see Ref. \cite{Calderon2013}).

\subsection{Basics of Kalman Filtering (KF)}
\label{sec:basicOfKF}

In this subsection, we review the key assumptions underlying the classic KF at a high level since the MBF makes several modifications to
this established algorithm \cite{hamilton,anderson1979}.  
  The KF assumes that a linear dynamical system can be used to describe the evolution of the ``state" (the ``state'' is 
the molecular position $r_{t_{i}}$ in our application and is not directly observable due to the measurement noise). 
The 
``process noise'' (i.e., the Brownian noise in Eq. \ref{eq:discreteSDE}) and net measurement noise are all assumed to be governed by Gaussian statistics in the classic KF.   
The KF leverages the following mathematical principle. Assume the random vector, 
$\left[\begin{smallmatrix}X \\ Y \end{smallmatrix} \right] \sim \mathcal{N}(\vec{\mu},\Sigma)$ where the mean, $\vec{\mu}=\left[ \begin{smallmatrix} \mu_X \\ \mu_Y \end{smallmatrix} \right]$, and covariance, $\Sigma = \left[ \begin{smallmatrix} \Sigma_{XX} & \Sigma_{XY} \\ \Sigma_{XY} & \Sigma_{YY} \end{smallmatrix} \right]$, specifying the Gaussian are known; $\Sigma_{XX}$  is the variance of $X$ (similarly for $\Sigma_{YY}$) and $\Sigma_{XY} := \mathrm{cov}(X,Y)$ is the covariance of $X$ and $Y$.  Assume $Y$ is directly observable, but $X$ is not.
In this case, the linear minimum variance  estimate of $X$ given $Y$, denoted by  $\mathbb{E}^*[X |Y]$ is given by \cite{anderson1979}:

\begin{align}
\mathbb{E}^*[X |Y] = & \mu_X + \Sigma_{XY} \Sigma_{YY}^{-1} (Y-\mu_Y ) 
\label{eq:optlsq}
\end{align}

\begin{figure}[htb]
\begin{minipage}[b]{1.\linewidth}
\def\pw{1}
% \begin{overpic}[width=\pw\textwidth]{./FigKFandMBF.pdf}
\begin{overpic}[width=\pw\textwidth]{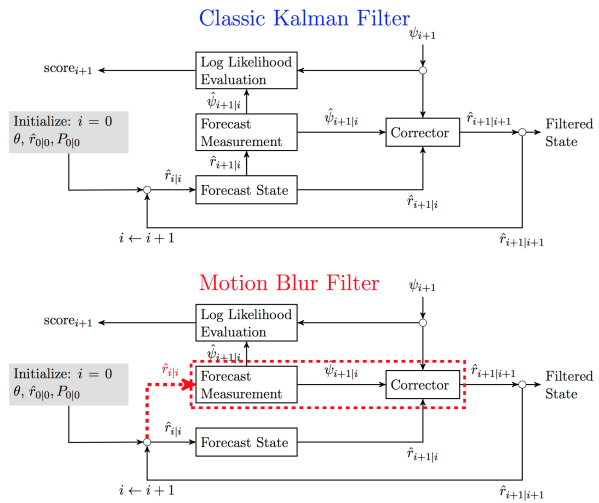}
\end{overpic}
\end{minipage}
% \begin{minipage}[b]{1.0\linewidth}
% \def\pw{1}
% \begin{overpic}[width=\pw\textwidth]{./Fig2KF.png}
% \end{overpic}
% \end{minipage}
% \begin{minipage}[b]{1.0\linewidth}
% \def\pw{1}
% \begin{overpic}[width=\pw\textwidth]{./Fig2MBF.png}
% \end{overpic}
% \end{minipage}
\caption{\footnotesize \textbf{Graphical Illustration of Kalman Filter (KF) and Motion Blur Filter (MBF).}  
 Each filter requires the parameters governing the process $\theta$ as well as the mean $\hat{r}_{0|0}$ and covariance matrix $P_{0|0}$ of the initial state as input.  Both filters  sequentially processes the measurements, $\psi_{i+1}$ to generate $\hat{r}_{i+1|i+1}$ which are estimates of the (unobservable) state $r_{i+1}$ at time $t_{i+1}$.  Two byproducts of each filter  are: (I) a likelihood score of the observation for the input $\theta$ (the sum of the logarithms of the likelihoods can be used to obtain the maximum likelihood estimate, $\hat{\theta}$ \cite{hamilton}) and  (II) summary statistics of both $r_{i+1}$ and $\psi_{i+1}$;  means are denoted by $\hat{r}_{i+1|i},\hat{\psi}_{i+1|i}$, respectively (the covariances are also computed, but omitted from the flow diagram to aid figure clarity).
 The top panel displays the classic discrete sequential KF and the bottom shows the new MBF.  The key algorithmic differences between the KF and MBF are highlighted by red dashed lines (see Sec. \ref{sec:methods} for mathematical details).  
}
\label{fig:flowChart}
\end{figure}

The above relationship is a general principle used by multiple estimators, not just the KF \cite{anderson1979}.  A useful aspect  of the KF algorithm is associated with the fact 
that a time series of measurements can be efficiently and sequentially processed building off of the general relationship in Eq. \ref{eq:optlsq}.  
To illustrate the sequential aspect and specialize to notation used in the discretized version of our model 
(Eqs. \ref{eq:discreteSDE} and \ref{eq:discreteMeas}), we use Fig. \ref{fig:flowChart} and the following equation:

\begin{align}
\label{eq:KFfilterUpdate}
\hat{r}_{i+1|i+1}    = & \hat{r}_{i+1|i} \ +  \mathrm{cov}(r_{i+1},\tilde{\psi}_{i+1})\mathrm{cov}(\tilde{\psi}_{i+1},\tilde{\psi}_{i+1})^{-1}
\tilde{\psi}_{i+1}, 
\end{align}

\noindent where  
 $\tilde{\psi}_{i+1} : = \psi_{i+1} - \hat{\psi}_{i+1|i}$, $\hat{\psi}_{i+1|i}$ is the ``Forecasted Measurement''  (the expected value of $\psi_{{i+1}}$) conditioned on the model parameters and all previously observed measurements up to time $t_{i}$ (the ``Forecasted State", $\hat{r}_{i+1|i}$, is analogously defined, but is used to predict  $r_{{i+1}}$). $\hat{r}_{i+1|i+1}$ represents the ``filtered state" estimate (i.e., the linear minimum variance estimate of $r_{{i+1}}$ given all information available up to time $t_{i+1}$).
 The sequence $\{ \tilde{\psi}_{i}\}_{i=1}^{T}$ is referred as the ``innovation sequence". Eq. \ref{eq:KFfilterUpdate} is the time sequential KF analog of Eq. \ref{eq:optlsq}.   The subscript 
 $t$ has been omitted from all quantities to simplify notation when dealing with discrete equations and filter algorithms (observing a subscript $i$ is equivalent to $t_{i}$).  
 The various boxes in Fig. \ref{fig:flowChart} compute different quantities in Eq. \ref{eq:KFfilterUpdate}.  For example,   the ``Forecast State'' box is one step of the algorithm and provides $\hat{r}_{i+1|i}$.  In the classic KF, the ``Forecast Measurement'' box  is an algorithmic step providing $\hat{\psi}_{i+1|i}$ given $\hat{r}_{i+1|i}$ (the corresponding MBF module uses a different input). 
The ``Corrector'' step of the algorithm combines the aforementioned forecasts and the actual measurement  $\psi_{i+1}$ to produce, 
$\hat{r}_{i+1|i+1}$. 
At the top of the diagram, we show that the forecasted measurements and observed measurement can be used to compute a likelihood score.  Beyond just providing an estimate  of the unobservable position $\hat{r}_{i+1|i+1}$, the KF can provide the MLE.  The MLE, $\hat{\theta}$,  is obtained by maximizes the sum of log likelihood scores associated with the observed $\{\psi_i\}_{i=1}^T$ over $\theta$ \cite{hamilton}.

Traditional KF measurement equations \cite{hamilton,anderson1979} often assume that an \emph{a priori} known linear transformation, $H$, which maps the state at the target filter time  of interest ($t_{i+1}$) to the measurement vector is available, i.e. $\psi_{i+1} = H r_{i+1} +$ ``measurement noise'' and the aforementioned measurement noise does not depend on past values of $r_s$ for $s < t_{i+1}$ \cite{hamilton,anderson1979}. Both conditions are violated for the motion blur model considered here and elsewhere \cite{Savin2005,Destainville2006,Berglund2010,Backlund2015}.  The time integral in Eq. \ref{eq:contMeas} is distributed as a Gaussian, but the mean of this measurement is in terms of $r_{i}$ vs. $r_{i+1}$ for measurement $\psi_{i+1}$. Also, the measurement noise in Eq. \ref{eq:discreteMeas} is statistically dependent on  $r_s$ for $s< t_{i+1}$.  More advanced treatments of the KF  show how to account for  ``Kronecker delta" type time correlations, i.e. $\delta_{ij}$, between  measurement and process noise (e.g., \cite{anderson1979}), however the time index offset shown in Eq. \ref{eq:discreteSDEcrossCorr} causes the KF technical challenges \cite{hamilton,anderson1979}.  The MBF overcomes these complications by fundamentally changing how $\hat{r}_{i|i}$  is  processed 
 (see  Fig. \ref{fig:flowChart}) as we outline in the next subsection.

\subsection{The MBF Algorithm}
\label{sec:methods}

In this section we provide the equations and
pseudocode outlining the MBF (recall the underlying discrete model processed by the MBF was presented in Eqs. \ref{eq:discreteSDE} - \ref{eq:discreteSDEnoCorr}).    
The ``parallel "processing of $\hat{r}_{i|i}$ by the Forecast Measurement and  Forecast State modules  shown in Fig. \ref{fig:flowChart} vs. the sequential processing used by the classic KF is one key aspect that distinguishes MBF from the KF.  The parallel processing of $\hat{r}_{i|i}$ is used because the time integral in Eq. \ref{eq:contMeas} violates core assumptions behind the KF \cite{hamilton,anderson1979} (discussed in previous section).  
Since the measurement noise induced by  motion blur  at $t_{i+1}$ depends on the process noise experienced in the interval $[t_{i},t_{i+1}]$, the MBF measurement forecast depends on linear transformations of  $\hat{r}_{i|i}$ and its covariance (vs. $\hat{r}_{i+1|i}$ and  its covariance  as in the classic KF).  

The parallel processing of $\hat{r}_{i|i}$ mentioned above requires some modifications to modules of  the classic KF algorithm shown in Fig. \ref{fig:flowChart}.  The modified ``Forecast Measurement" equation reads: 

\begin{align}
\hat{\psi}_{i+1|i} = & H_A + H_F\hat{r}_{i|i} \\
S = & H_FP_{i|i}H_F^\top + R_i
\end{align}

\noindent and the modified ``Corrector'' (also known as the  ``Measurement Update'' \cite{anderson1979}) equation reads:

 \begin{align}
 \label{eq:gainNew}
 K := & (C+FP_{i|i}H_F^\top)(H_FP_{i|i}H_F^\top+R_i)^{-1} \\
  \label{eq:filterUpdate}
 \hat{r}_{i+1|i+1} = & \hat{r}_{i+1|i} + K(\psi_{i+1} - \hat{\psi}_{i+1|i})  \\
 \label{eq:filterCovUpdate}
 P_{i+1|i+1} = & P_{i+1|i} -  K(H_FP_{i|i}H_F^\top+R_i)K^\top 
 \end{align}

\noindent where the filter state forecast ($\hat{r}_{i+1|i}$), the filtered state ($\hat{r}_{i+1|i+1}$),  and the measurement forecast ($\hat{\psi}_{i+1|i}$) described in the previous section have covariances  $P_{i+1|i}$, $P_{i+1|i+1}$ and $S$, respectively.  We use $S$ to denote the ``innovation covariance'' \cite{anderson1979} (the notation $P_{i+1|i}$ and $P_{i+1|i+1}$  is common in the KF framework  \cite{hamilton,anderson1979}).    
Recall that the parameters $A,C,F,H_A,H_F,Q,R_i$ are associated with the statistically precise discretized version of Eqs. \ref{eq:SDE} - \ref{eq:contMeas} and assume  $t_E = t_{i+1}-t_i$ for all $i$,
  (expressions for these parameters, depending on $t_E$, are provided in the previous section and Appendix).

Note that the altered measurement forecasts  affects the form of the classic ``Corrector'' or ``Measurement Update" equations  \cite{anderson1979}.
The measurement noise induced by motion blur is correlated to the process noise 
under the assumed model and this changes the standard form of the ``Corrector'' update (Eq. \ref{eq:KFfilterUpdate}), specifically under the MBF model:
\begin{align} 
\mathrm{cov}(r_{i+1},\psi_{i+1}) :=  \mathrm{cov}(r_{t_{i+1}},\psi_{t_{i+1}}) \\
 =  \mathrm{cov}(A+Fr_{t_i} + \eta_{t_{i}},H_A+H_Fr_{t_i} + \epsilon^{\mathrm{loc}}_{t_{i+1}} + \epsilon^{\mathrm{mblur}}_{t_{i+1}}) 
 \label{eq:crossCor}
\end{align} 

\noindent As shown in the Appendix, 
$C := \mathrm{cov}(\eta_{t_{i}},\epsilon^{\mathrm{mblur}}_{t_{i+1}}) \ne 0$.  The correlation relationship in  Eq. \ref{eq:crossCor} is used to alter the form of the standard ``Corrector'' or ``Measurement Update"  and the  end result was shown in Eq. \ref{eq:gainNew} above (see pgs. 116-117 in Ref. \cite{anderson1979} for complete theoretical background).
The remaining equations defining the classic KF are the same in the MBF.  For example, since 
 $\mathrm{cov}(\eta_{t_{i+1}},\epsilon^{\mathrm{mblur}}_{t_{i+1}}) = 0$  within the model considered, the ``Forecast State'' updates,
  $\hat{r}_{i+1|i} =A+F\hat{r}_{i|i} $ and $P_{i+1|i} = FP_{i|i}F^\top + Q$, associated with the classic KF are still valid \cite{anderson1979,hamilton}.

Pseudocode implementing the theoretical ideas above is provided in Algorithm \ref{mainAlg}.  Note that the program flow was set up so that the classic KF could also be implemented  within the MBF' algorithmic framework.   Eq. \ref{eq:gainNew} has a fairly different form in the MBF compared to the KF.   This change was required due to the aforementioned parallel processing of $\hat{r}_{i|i}$ illustrated in Fig. \ref{fig:flowChart} 
(changing other equations from the MBF to the KF essentially requires changing $P_{i|i}$ to $P_{i+1|i}$ as shown in the code provided on GitHub).   One difference worth noting is that $H_F$ and $H_A$ have different definitions when Algorithm \ref{mainAlg} is used to process the classic KF (the open-source code provided illustrates this feature). 
 It should also be noted that although we focus on the 1D scalar case in this article,  the MBF algorithm presented above is described in terms of the multivariate case.

\SetNlSty{normal}{}{   } 
\begin{algorithm}
    \SetKwInOut{Input}{Input}
    \SetKwInOut{Output}{Output} 
    
    function  \url{MotionBlurFilter} $(\{\psi_i\}_{i=1}^T, \{\sigma_{\mathrm{loc}}^{{\mathrm{Input}}}(i)\}_{i=1}^T, \theta )$ \newline
 
    \% \textbf{Inputs}: Time series of noisy position measurements $\{\psi_i\}_{i=1}^T$, candidate parameter vector $\theta = (\kappa,D,\sigma_{\mathrm{loc}},v)$, and [optional: time series of localization estimates,$\{\sigma_{\mathrm{loc}}^{{\mathrm{Input}}}(i)\}_{i=1}^T$]  \newline
    \% \textbf{Outputs}: log likelihood $\log \mathcal{L}$ and filtered state series $\{r_{t|t}\}_{i=0}^T$ \newline

\%Compute Discrete Filter Variables  

    $P_{1|0},r_{1|0},r_{0|0},P_{\mathrm{Innov}}=$  \url{InitializeFilterPars} $(\{\psi_i\}_{i=1}^T,\theta,\delta)$ \\
    $F,Q,A,H,Q^{\mathrm{mblur}},C,H_F,H_A =$  \url{ExactMapOfContinuousToDiscrete}($\theta$)
\%Note: some auxiliary variables above are solely to allow this routine to process the classical form of the Kalman Filter. \label{alg1:contToDiscrete} \newline

    filteredState={$r_{0|0}$}; $\log \mathcal{L}$ = 0 \% Initialize variables to be returned  \newline

    \For{$t = 1:T$}{
$R_t$ = $Q^{\mathrm{mblur}}$ + $(\sigma_{\mathrm{loc}}^{{\mathrm{Input}}}(t) + \sigma_{\mathrm{loc}})^2$  \%Compute Net Measurement Noise Covariance at $t$ \label{alg1:Rupdate}  \newline

\% Begin computation of innovation likelihood \\
$S = H P_{\mathrm{Innov}} H^\top + R_t$ \ \ \%Compute Innovation Covariance at $t$  \label{alg1:evalMeasUncertainty}  \\
$z = \sqrt{S^{-1}}(\psi_t-H_F r_{t-1|t-1}-H_A)$ \ \ \%Normalized Innovation at $t$ \label{alg1:forecastMeas}  \\
$\log \mathcal{L} = \log \mathcal{L} + 1/2 \log\Big(|(2\pi S)^{-1}|\Big)  -z^\top z/2$ \ \ \%add to log likelihood \newline

\% Update filter parameters for next iteration \\
$K =$ \url{ComputeGain}($P_{\mathrm{Innov}},C,H,R_t,F$) \label{alg1:computeGain}  \\
$r_{t|t} = r_{t|t-1} + K (\psi_t-H_Fr_{t-1|t-1}-H_A)$ \\
filteredState.append($r_{t|t}$) \%Store filter estimate \\

$P_{t|t} = P_{t|t-1} -  K(HP_{\mathrm{Innov}}H^\top+R_t)K^\top$ \%Update filter covariance \label{alg1:updateFilterState}  \\
$P_{t+1|t}  = FP_{t|t}F^\top + Q$ \%Update state forecast covariance \label{alg1:forecastStateCov}  \\
$r_{t+1|t}=Fr_{t|t} + A$ \%Forecast state \label{alg1:forecastState}  \\
$P_{\mathrm{Innov}}$ =  \url{ComputeInnovCov}($P_{t|t},P_{t+1|t}$) \label{alg1:switchInnov}  \\
\% Call above allows Algorithm to also process classic KF \\

} 

\textbf{return} \ \ $\log \mathcal{L}$,filteredState

\caption{Pseudocode for evaluating the innovation likelihood of common SPT models (pure, directed, and confined diffusion) given time series of correlated observations obscured by motion blur and localization  measurement errors.  External functions appearing below are flagged via a different font  and are defined in the Supp. Mat.}
\label{mainAlg}
\end{algorithm}

\subsection{Illustrative Results and Introduction to Software}
\label{sec:qualitativeExamples}
To show the modeling ideas applied to practical SPT problems, we illustrate how the MBF output can process two types of  trajectories commonly encountered in SPT where other analysis methods encounter problems (these trajectories exhibit some form of statistical non-stationarity).   In these simulations, we generate exact realizations from the OU process shown in Eq. \ref{eq:SDE}.  The IPython Notebooks supplementing this work have comments describing the input parameters. We describe the basic simulation at a high-level in the next paragraph and the relevance of the results afforded by the algorithm \& software tool introduced within this article in the remaining paragraphs.  Sec. \ref{sec:results} presents larger scale simulations  results (studying multiple  trajectories  under  systematically varied parameter regimes).

Before proceeding, we need to introduce some simulation parameters: $N_{\mathrm{sub}}$ is a parameter used to model the underlying particle visiting multiple spatial locations while emitting photons used to construct a PSF in one image. 
To model the microscope's measurement output $\psi_{t_i}$, the OU trajectory was sampled at $t_i - (\frac{N_{\mathrm{sub}} - 1}{N_{\mathrm{sub} } })t_E, t_i - (\frac{N_{\mathrm{sub}} - 2}{N_{\mathrm{sub} } })t_E, \ldots, t_i $ and then averaged over the $N_{\mathrm{sub}}$ samples to mimic ``motion blur'' in one frame.  The effective localization errors induced by finite photon counts and background fluorescence on the  motion blurred trajectory is modeled by a mean zero Gaussian, $\epsilon^{\mathrm{loc}}_{t_i}$, and this Gaussian random variable (independent of $r$) is added to the discretely sampled and ``blurred'' trajectory.  This process for simulating measurements is repeated for each of the uniformly spaced observation time.  We subsequently attempted to infer/extract the parameter $\theta$ given one trajectory generated in this fashion with various estimators.   IPython  
(Jupyter) Notebooks generating the data and graphs are provided to facilitate users implementing these techniques.
Additional simulation details are deferred to these notebooks since we focus on illustrating new capabilities in this section.

\begin{figure}[htp]
\centering

\begin{minipage}[b]{.85\linewidth}
\def\pw{1}
% \begin{overpic}[width=\pw\textwidth]{./Fig2b.pdf}
% % \put(38,79){\huge Time [s]}
% % \put(33,50){\huge Time Lag [s]}
% \put(36,79){\large Time [s]}
% \put(32,50){\Large Time Lag [s]}
\begin{overpic}[width=\pw\textwidth]{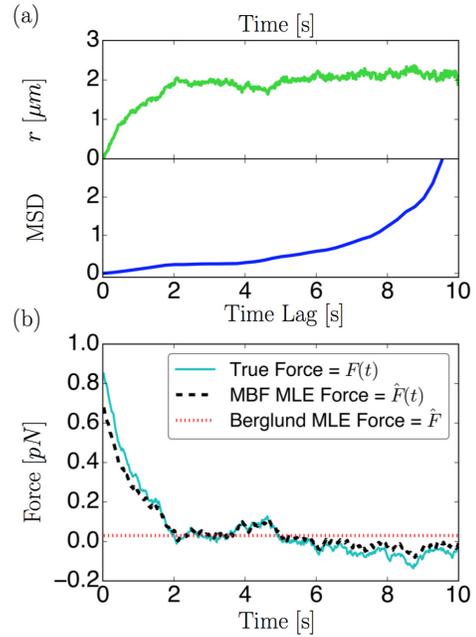}
\end{overpic}
\end{minipage}

\caption{\footnotesize \textbf{Illustration of Issues Encountered by Established SPT Methods.}
The top panel displays the measured position of a simulated non-stationary confined trajectory. The middle panel displays the corresponding Mean Squared Displacement (MSD) vs. time lag.  The bottom panel displays the true instantaneous force of the simulated particle as well as two estimates of the time dependent force (i) that of a ``directed diffusion'' model accounting for motion blur, but assuming a constant velocity over time and (ii) that of the Motion Blur Filter (MBF) introduced here (see text for details).
   Recall that position and force are quantities defined instantaneously in time for a single trajectory. This is in contrast to the information contained  at single ``time lag'' value (denoted by $\tau$) in the MSD vs. time lag curve. MSD curves  empirically average the square of position differences $\big(r(t+\tau)-r(t)\big)^2$ over multiple time windows where $t$ represents absolute time (the MSD curve is obtained by varying $\tau$). IPython Notebooks used to generate these figures are provided with this work to provide additional simulation and 
parameter estimation details.
}
\label{fig:nonstatVelocityEG}
\end{figure}

The first example analyzes a trajectory where non-stationary phenomena affect the dynamics of the observed measurement sequence.  In the trajectory shown in Fig. \ref{fig:nonstatVelocityEG}, the particle is being ``sucked into a harmonic well''.  There are large attractive forces at earlier times and as this particle relaxes into the harmonic well, these forces reduce in magnitude (the forces are precisely quantified in the bottom panel of Fig. \ref{fig:nonstatVelocityEG}).    
The molecule's mean position and ``position increments" (the latter is used in MSD) also change appreciably over time (i.e., neither the position or ``confinement forces'' have reached their stationary distribution \cite{risken}). 
This relaxation induces the primary source of statistical 
non-stationarity in this example.  The true instantaneous velocity  in these simulations can be obtained precisely via evaluating  $v(t_i) = v -\kappa r_{t_i}$. The force, $F(t_i)$, is obtained by dividing the velocity by $\frac{D}{k_BT}$ at each point.  We used the noisy position vs. time data, $\{ \psi_{t_{i}}\}_{i=1}^T$, to infer the time dependent force by using $\{ \psi_{t_{i}}\}_{i=1}^T$ and the MLE computed by the MBF algorithm.  

Note that a stationarity assumption is often implicit in MSD or autocorrelation (including Fourier transform) based approaches \cite{Savin2005,Destainville2006,Backlund2015}.    The MSD (computed with the 400 samples) is shown in Fig. \ref{fig:nonstatVelocityEG} and
illustrates artifacts induced by the fact that the position increment distribution changes over time.  
    The MBF's estimate  of the instantaneous  velocity
 is $\hat{v} -\hat{\kappa} \hat{r_{i|i}}$ (force is estimated by dividing this by $\frac{\hat{D}}{k_BT}$).
 Recall that, data-driven MLE parameters are denoted by hats and $\hat{r}_{i|i}$ denotes the MBF's estimate (using the MLE) at the underlying position given the measurements up to time $t_i$ (the true position is not observable due to static and dynamic errors).  It is emphasized that we do not use a finite difference (FD) scheme to estimate velocity (i.e., a FD scheme takes differences of measurements and divides by the time between observations) since realized SDE paths are not mathematically differentiable \cite{kp} Though the ``drift function'' (terms in front of $dt$) of the model can provide a mathematically well-defined  ``instantaneous velocity'' \cite{kp}.  The Supp. Mat. provides an illustration of the output of a simple FD scheme applied to this trajectory to highlight this problem.  Fig. \ref{fig:nonstatVelocityEG} also displays a 
 modified  Berglund \cite{Berglund2010} algorithm (accounting for constant velocity)  estimate of average force.  With data sampled at 25$ms$ for $T=400$ observations,  one can obtain an accurate trajectory of both velocity and force.  Note that in the parameterization shown 
 in Eq. \ref{eq:SDE}, the drift function provides a model of the instantaneous velocity (the velocity plot is shown in the Supp. Mat. Fig. 5).  Extracting the instantaneous force from the drift function requires appealing to a fluctuation dissipation relationship as well as an accurate estimation of the diffusion coefficient.  It is envisioned that there are many situations where one might not want a time averaged force or velocity (e.g., one would miss the ``relaxation event'' at earlier times of this trajectory).   Furthermore, as we show in Fig. \ref{fig:diffProfile}, the MBF allows unbiased estimation of the diffusion coefficient, $D$ over a wide range of molecular diffusivities and exposure times, whereas other state-of-the-art methods introduce biases in $D$ (and hence biases in estimated force). 
 
 \begin{figure}[htb]
 \centering
\begin{minipage}[b]{.85\linewidth}
\def\pw{1}
% \begin{overpic}[width=\pw\textwidth]{./Fig3bNew.pdf}  
% % \put(38,98){\huge Time [s]}
% % \put(34,50){\huge Time Lag [s]}
% \put(36,98){\large Time [s]}
% \put(32,50){\Large Time Lag [s]}
\begin{overpic}[width=\pw\textwidth]{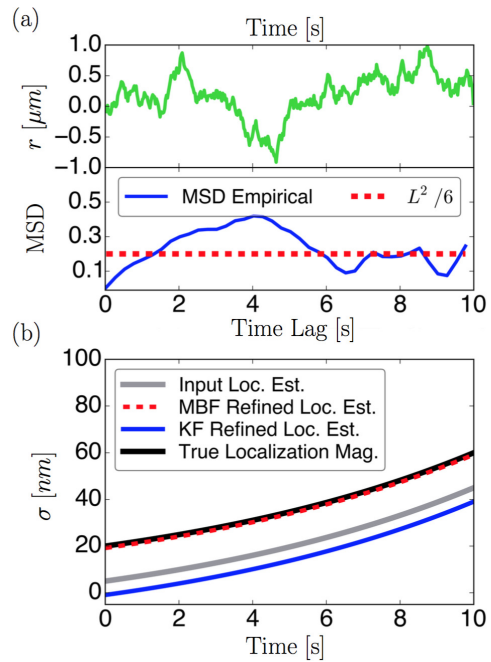}
\end{overpic}
\end{minipage}
\caption{ \footnotesize \textbf{Empirically Estimating a Non-stationary Effective Static Noise.}
The top panel displays the measured position of a simulated confined trajectory experiencing a localization noise variance changing over time. The middle panel displays the corresponding Mean Square Displacement (MSD) curve.  Note the top and bottom plots have ``absolute time'' for the x-axis and the MSD plot uses ``Time Lag'' for the x-axis (see caption of Fig. \ref{fig:nonstatVelocityEG} for additional details regarding the MSD curve).
The bottom panel displays (i) the magnitude of the true time varying localization noise added to this trajectory, (ii) a nominal ``Cramer-Rao'' type lower bound proxy of the localization noise provided to the time series estimator (labeled as ``Input Loc. Est.''), (iii) two data driven-estimates of the true localization noise.  One estimates uses
the classic Kalman Filter (KF) and the other uses the Motion Blur Filter (MBF) to estimate  the effective  localization noise (note that motion blur in the measurements causes the KF to be misspecified).
IPython Notebooks used to generate these figures are provided with this work to provide additional simulation and parameter estimation details. 
}
\label{fig:RvaryingEG}
\end{figure}

The second example focuses on how our approach can utilize statistics characterizing localization uncertainty information computed in individual  images. The top 
panel of Fig. \ref{fig:RvaryingEG} displays a confined trajectory.  
In this trajectory, the effective localization noise  is simulated to increase over time due to photobleaching effects (introducing non-stationarity in measurement statistics). Time dependent localization noise is commonly encountered when multiple GFP dyes are used to tag a molecule and/or background, e.g. \cite{Thompson2010,Calderon2013b}.  
The nominal localization noise in the input sequence $\{\sigma_{\mathrm{loc}}^{{\mathrm{Input}}}(t_i)\}_{i=1}^T$ is intended to come from a typical SPT localization method applied to a PSF (e.g., \cite{Thompson2002,Ober2004}), however it is acknowledged that the PSF measured not only
contains contributions from static and dynamic errors, but the uncertainty estimate is likely an idealized limit (hence the data-driven effective static measurement noise magnitude will differ from the input sequence).  The version of the sample software provided permits estimation of a constant off-set, $\sigma_{\mathrm{loc}}$, from the input $\{\sigma_{\mathrm{loc}}^{{\mathrm{Input}}}(t_i)\}_{i=1}^T$ (the code can be easily modified to account for more complex models). If no input sequence is provided, the code estimates a constant time independent effective static measurement noise.

The (known) true static localization error magnitude of the noise added to the trajectory is shown in the bottom panel of Fig. \ref{fig:RvaryingEG}.  Parametric or non-parametric estimates of this type of noise trend from the time series data alone can be difficult if trajectories are not long \cite{Calderon2013b} and/or if the exposure times associated with the measurements are large.  
Fortunately, various theoretical approximations for the lower-bound of the uncertainty associated each localization at $t_i$
can be obtained with established methods \cite{Thompson2002,Ober2004,Abraham2010}.  However, these uncertainty estimates often appeal to large sample Cramer-Rao bounds (CRB) \cite{Ober2004,Abraham2010,Thompson2010} which are not reflective of the true ``static error'' observed in practice. Finite sample error and real-world features (including motion blur) not accurately modeled often make the empirical data exhibit error different than the CRB.  The bottom panel of Fig. \ref{fig:RvaryingEG} displays a time varying CRB type localization estimate.  Even though the CRB estimates are overly optimistic,
this type of time dependent localization information (i.e., uncertainty estimates provided by a 3rd party piece of software) can be  used to aid kinetic analysis.

  Both the classic KF and the MBF are able to  utilize the noisily measured data 
    and the (possibly biased) time dependent localization information afforded by image analysis \cite{Ober2004,Abraham2010,Thompson2010}   to jointly infer the kinetic parameters.  Recall that the MBF code provided estimates a constant off-set adjustment $\sigma_{\mathrm{loc}}$ to the input localization noise standard deviation (an estimated zero off-set of the localization noise implies perfect agreement with the input static noise standard deviation).  The MBF can use the observations and blur information encoded in Eqs. \ref{eq:SDE}- \ref{eq:contMeas} to consistently estimate both the empirical effective static measurement noise and the kinetic parameters governing motion despite slightly biased time varying localization input. The KF assumes the measurements are reflective of the instantaneous position of the particle at the time of the measurement and  underestimates the net static noise magnitude (this effect is expected from the results reported originally in Ref. \cite{Berglund2010} for a ``pure'' diffusion case).  
  
  It is emphasized throughout that accurately modeling fluctuations and measurement noise enables one to extract higher quality information from individual trajectories.  
  For example, $L^2/6$  (the long time MSD limit  assuming the SDE parameters are fixed and $\kappa>0$ \cite{Kusumi1993,Park2010,Calderon2013}) is accurately estimated via the MBF.  An MLE estimate of 0.22 was obtained from the data using a single  trajectory (400 observations spaced by $25ms$) without requiring the selection of tunable parameters (the true value of  $L^2/6$ was 0.20). 
In  MSD computations processing a single trajectory,  the ``time lag'' is often denoted by $\tau$ and a single point in an MSD curve fixes $\tau$ and computes the average of $\big(r(t+\tau)-r(t)\big)^2$ over time.
  However, many estimates relying on the MSD  include biases induced by ``time lag truncation''.  
  In practice, the full trajectory is rarely used for MSD computations \cite{Thompson2010,Chen2014} since there is less data available for larger values of $\tau$. Using a subset of the available $\tau$ values  is what we refer to as ``time lag truncation'' error.  
  Time lag truncation is usually carried out heuristically  and the decision of where to truncate can affect parameter estimates based on MSD \cite{Berglund2010,Thompson2010}.
   For example, if one averages over the last 1/4 of the MSD displayed in Fig. \ref{fig:RvaryingEG}, an estimate of
$L^2/6 = $   0.17 is obtained for the plateau value.  If one ignores the first and last quarters of the data, an estimate of 0.28 is obtained (such heuristic truncations are commonly used in MSD and this is a well-known problem with MSD \cite{Berglund2010,Thompson2010}). 
  Note that MSD curves also inherently include 
  unnecessary additional noise source (e.g.,  noise due to aggregating position increments over disparate times).  
Our approach, using the full sequence of data (without tunable parameters), is close to the truth despite the high degree of noise observed in the MSD.   
The ability to systematically leverage  time dependent localization information (afforded by physics based models \cite{Thompson2002,Ober2004,Abraham2010,Gahlmann2013,Liu2015}) into the MBF and carry out likelihood inference is practical benefit of the MBF approach. Hence, we have provided examples of how to achieve this in our associated IPython Notebooks.

\subsection{Comparison to Other Approaches}
\label{sec:compToOtherApproaches}
A variety of techniques have attempted to utilize MSD approaches to quantify both static and dynamic error statistics \cite{Savin2005,Destainville2006,Backlund2015}. However, as we illustrated in the previous subsection, MSD approaches ignore useful time-ordered information.  Specifically MSD methods aggregate increments from potentially disparate times. This aggregation can  degrade dynamic information and complicate analyzing the MSD.  Likelihood based techniques have been applied to SPT tracking problems previously \cite{Montiel2006,Voisinne2010,Hoze2014,Beheiry2015}, though,  the aforementioned works ignore the time correlation effects induced by purely static error  in addition to making unnecessary approximations of the likelihood function.  One approximation used is the so-called Euler (sometimes referred to as the  Euler–Maruyama \cite{kp}) approximation.  The Euler approximation is a numerical integration technique which simulates a generic SDE, $dr_t=\mu(r_t)dt + \sigma(r_t)dB_t$, via $r_{t_{i+1}} = \mu(r_{t_{i}})\Delta t_{i} + \sigma(r_{t_{i}})\Delta B_{t_{i}}$ where $\Delta t_{i} := t_{i+1}-t_{i}$ and $\Delta B_{t_{i}}:= B_{t_{i+1}} -B_{t_{i}}$ (this Brownian increment can be simulated precisely, but the other numerical approximation errors can be large \cite{kp,Jensen2002}). 
We remind the reader  that  Eq. \ref{eq:discreteSDE} - \ref{eq:discreteMeas} solve the assumed SDE and measurement equation precisely without any numerical integration or Euler type approximation (hence temporal and spatial statistics are consistent with the SDE model in Eq. \ref{eq:SDE} - \ref{eq:contMeas}).
Time series methods appealing to the Euler approximation   can cause a high degree of parameter estimation bias even in the measurement noise free cases \cite{Jensen2002,llglassy}.  
Ignoring statistical time correlation induced by localization and motion blur further degrades estimates of parameters.  Inaccurate likelihood approximations (like those induced by the Euler approximation) also prevent  researchers from applying reliable consistency tests to fitted models since the likelihood does not correspond to the assumed model \cite{llglassy}. 

In addition to the aforementioned issues, implicit spatial or temporal stationarity assumptions are made in many SPT approaches \cite{Voisinne2010,Vestergaard2014,Hoze2014,Beheiry2015}.
The first work (to this author's knowledge)  treating static and dynamic error induced by motion blur in SPT using a likelihood based approach 
 was 
Berglund's pioneering work \cite{Berglund2010}. 
 Berglund \cite{Berglund2010} considered a constant diffusion model contaminated by static and dynamic error 
(extending to a ``directed'' or constant velocity model, where velocity is time and space independent, is straightforward due to the measurement difference formulation used as shown by the Python companion code).  However, measurement difference \cite{Savin2005,Destainville2006,Berglund2010,Backlund2015} based schemes   typically make a time stationarity assumption. That is, they assume that moments and  time correlations of increments of measurements,  $\psi_{t+\tau}-\psi_{t}$ are independent of $t$.   Stationary assumptions are also commonly made in power spectral methods \cite{Vestergaard2014} and ``nonparametric'' approaches \cite{Hoze2014}. 
In Refs. \cite{Voisinne2010,Beheiry2015}, an approximate   Bayesian approach was used to approximate temporally and spatially  dependent velocity and force (an Euler approximation of the likelihood was utilized), but  the approach did not account for the time correlation effects of static or dynamic measurement noise (i.e., established KF ideas were not used and the likelihood was inexact).
Maximum likelihood time series estimation of parameters determining spatially dependent velocity and force using the standard KF likelihood have been studied   in single-molecule manipulation studies \cite{SPAfilter,Fernandez09,SPAfric,SPAdsDNA} and in SPT \cite{Calderon2013b,arxivDec2013}. Ref. \cite{CalderonBloom2015} extended the KF to allow ``switching linear dynamical systems'' and used a nonparametric Bayesian approach to systematically determine regime switching with an exact likelihood.
However, in Refs. \cite{Calderon2013b,arxivDec2013,CalderonBloom2015}, effects of motion blur were lumped into the effective measurement error since both $D$ and $\delta$ were small and the ``exact" KF likelihood was computed corresponding to a model only technically accounting for static measurement errors.

\section{Results and Discussion}
\label{sec:results}
In the results that follow, we focus on analyzing $N$ time series containing $T$ observations with a uniform time spacing, $\delta$, between observations. The data is modeled as being collected in uniform continuous illumination, a common situation in cell biology \cite{Berglund2010,Gahlmann2013,Liu2015}.  For all simulations reported in this section, we study a constant localization noise, $\sigma_{\mathrm{loc}} = 30 \ nm$, $v=0$, $\kappa > 0$, and initial conditions drawn from the stationary distribution in order to focus on the  effects of motion blur on confined trajectories.   To facilitate comparison and reduce noise due simply to random number generation, we analyze the same batch of trajectories with three estimators: the modified Berglund directed diffusion model with motion blur, the classic KF (without motion blur), and the new MBF.
To illustrate that our derived variance and mean formulas are valid, we simulate $N_{\mathrm{sub}}=100$ points spaced by 
$\frac{\delta}{ N_{\mathrm{sub}} }$ using the exact known solution to the OU process and average these quantities to approximate the integral in the  Eq. \ref{eq:contMeas}. The $N_{\mathrm{sub}}$ parameter determines the accuracy of approximating the integral in Eq. \ref{eq:SDE}. We did  not simulate the discrete realizations from the analytically derived motion blur results (derived in Appendix) to illustrate that our equations are correct and perform reasonably even if the  uniform continuous illumination model contains discretization errors. In practice, pixelation and other factors \cite{Abraham2010} often introduce discrete sampling errors not completely captured by the  motion blur model in Eq. \ref{eq:contMeas}. 
   Each parameter estimate reported in this section used $N=T=400$ and a separate MLE was obtained for each of the $N$ trajectories (i.e., trajectories were not combined to find a single parameter vector). The Supp. Mat. reports results with shorter trajectories  corresponding to the main plots shown ($T=100,N=400$).

 \begin{figure}[htp]
\centering
\begin{minipage}[b]{1.0\linewidth}
\def\pw{1}
% \begin{overpic}[width=\pw\textwidth]{./Fig4_DpanelNew2}
\begin{overpic}[width=\pw\textwidth]{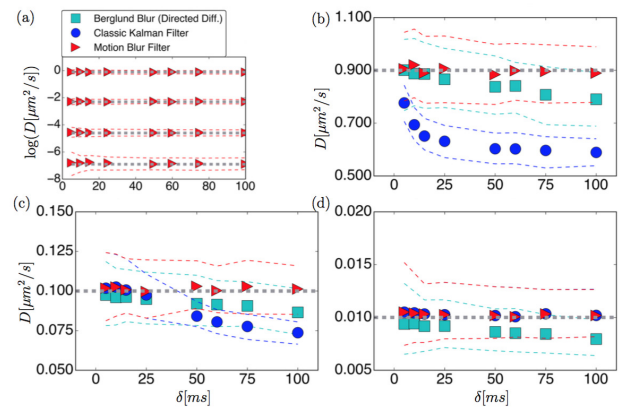}
\end{overpic}
\end{minipage}
\caption{\footnotesize \textbf{Demonstration of MBF Advantages when Estimating Diffusion Coefficients.}
Diffusion coefficient estimation results for $\kappa=1 s^{-1}, \sigma_{\mathrm{loc}} = 30 nm, \ v=0$. Recall $\delta$ determines the temporal resolution of the measurement (as $\delta$ increases, both motion blur and confinement effects become pronounced).  The upper left panel plots the median (symbols) and the 10th / 90th percentiles (dashed lines) of the Motion Blur Filter MLE parameter distribution computed using $N=400$ trajectories of length $T=400$ trajectories (each trajectory produced an MLE estimate) on log scale for  various $D's$ and $\delta$'s (each value of $D$ and $\delta$ corresponds to a summary of $N$ simulations of length $T$).  The remaining panels zoom in on the larger $D$ cases and show results obtained by applying other estimators to the same collection of trajectories.  The other likelihood based estimators fail for different reasons.  Berglund's blur model \cite{Berglund2010} fails due to confinement effects becoming more pronounced as $\delta$ increases. The Kalman Filter handles confinement when motion blur effects are low (e.g., see bottom left panel), but fails when motion blur is amplified (motion blur effects increase with both $\delta$ and $D$).  This result demonstrates the MBF is robust to a variety of regimes of relevance to SPT data modeling.}
\label{fig:diffProfile}
\end{figure}

In Fig. \ref{fig:diffProfile}, we fix $\kappa= 1 s^{-1}$, $v=0$, and analyze various diffusion coefficients
$D = 1\times10^{-3},1\times10^{-2},1\times10^{-1},0.9\times10^{-1} \mu m^2/s$ (the latter two values were inspired by the lattice light sheet MSD results reported in Ref. \cite{Chen2014}) and $\delta$'s ranging from 5$ms$ to 100 $ms$.  In the upper left panel, we plot the median (solid symbol) and $10^{th}$ and $90^{th}$ percentiles (dashed lines) of the empirical parameter distribution obtained by analyzing the $N$ trajectories and obtaining the MLE of the MBF on logarithmic scale.  The other panels zoom in on the diffusion coefficient estimates and also show the modified Berglund estimator (this estimator models motion blur and constant velocity, but ignores spatial variation in velocity) and the classic KF estimates (the KF estimator ignores motion blur, but models spatial variations in velocity).  Note that the MBF consistently estimates the diffusion coefficient over the wide range of $D$'s' and $\delta$'s considered.  The other two estimators fail for different reasons.  For large $D$   and/or large $\delta$ (where large is relative to typical SPT studies), the effects of motion blur become pronounced and bias  the KF's diffusion coefficient estimation.  Note that for smaller $D$ (consistent with large macromolecular complexes), $\delta$ must be quite large before unmodeled motion blur affects diffusion estimation.  In the Berglund estimator, the effects of nonzero $\kappa$ inducing confinement become more pronounced at larger $\delta$ and this adversely affects $D$ estimation regardless of the magnitude of $D$ (despite motion blur being modeled in this model).  \emph{The ability of  the MBF algorithm to reliably model $D$'s relevant to SPT in the presence of varying degrees of confinement  and exposure times is expected to aid researchers in SPT.}

 \begin{figure}[htp]
\centering
\begin{minipage}[b]{1.0\linewidth}
\def\pw{1}
% \begin{overpic}[width=\pw\textwidth]{./Fig5.pdf}
\begin{overpic}[width=\pw\textwidth]{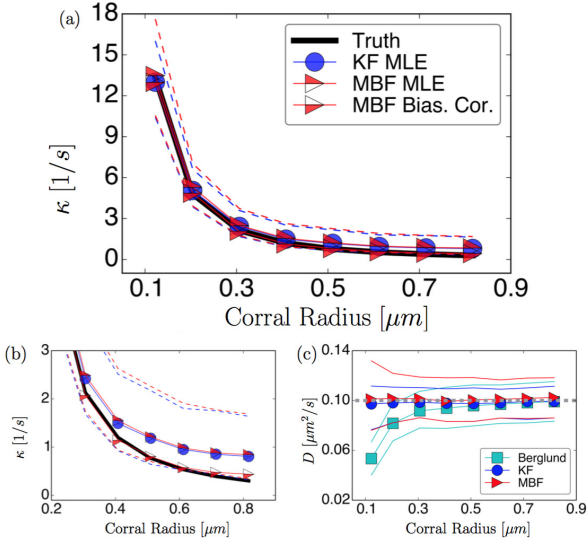}
\end{overpic}
\end{minipage}
\caption{\footnotesize \textbf{Stability of the MBF in Estimating Kinetic Parameters Under Various Degrees of Confinement.} Estimation of various $\kappa$ for fixed $D=0.1 \mu^2 m/s, \   
 \sigma_{\mathrm{loc}} = 30 nm, \ \delta = 25 ms, \ v=0$.  Similar to Fig. \ref{fig:diffProfile} in terms of $N$ and $T$, except different kinetic parameters were varied.   In all panels, the $x-$axis displays the so-called ``corral radius"  $:= \sqrt{\frac{L^2}{6}}$ (see Sec. \ref{sec:physicalSDE}). In the top panel, the $y$-axis displays the median $\hat{\kappa}$  (symbols) and the 10th / 90th percentiles (dashed lines) of the estimated Maximum Likelihood Estimate (MLE) parameter distribution for the corral radii explored.  The bottom left panel zooms in on the ``less confined" cases.  $\kappa$ estimated with the Kalman Filter (KF) and Motion Blur Filter (MBF) were similar.  The results of a finite $T$ bias correction \cite{Calderon2013} are also displayed (this correction is fairly close to the known true value).  The bottom right panel shows the corresponding estimates of $D$ for the same trajectories and estimators considered.
 }
\label{fig:kappaProfile}
\end{figure}

Next we fix $D=0.1 \mu m^2/s$ and  $\delta = 25 ms$ and vary $\kappa$ in Fig. \ref{fig:kappaProfile} (the Supp. Mat reports  $\delta = 10 ms$ results).  This value of $D$ was selected since the classic KF and MBF were shown to be similar for this range of values.  The interest is in determining the accuracy of the estimated $D$ and $\kappa$ as the latter varies in the presence of motion blur.  A large positive value for $\kappa$ corresponds to high confinement or a small ``corral radius".  
Recall that the ``corral radius" $:= \sqrt{\frac{L^2}{6}} = \sqrt{\frac{2D}{\kappa}}$ and the parameter $L$ quantifies the length of the region the particle can explore under confinement (see Sec. \ref{sec:physicalSDE}).  At small  $\delta$, estimates of $D$  are consistent with one another for both the classic KF and the new MBF, however the Berglund motion blur MLEs only begins to converge to the other two as the corral radius is increased (i.e., as confinement decreases).  We point out that the rate of convergence of the Berglund $D$ estimate to that of the MBF (the MBF nests the Berglund estimator considered)  is primarily dictated by $\delta$ for a fixed $\kappa$.  

In Fig. \ref{fig:kappaProfile}, it is shown that estimates of $\kappa$ obtained using the KF and MBF likelihoods  are relatively close to one another for the corral radius values studied  ($\kappa$ is not reported for the Berglund estimator since it is not included in this model).  
However, the median of the KF and MBF's MLE  are biased from the known truth due to the discretely sampled finite length  trajectories producing the MLE vector (the MLE mean/median converges to the truth as $T\rightarrow \infty$ for a correctly specified OU model).  This finite time series sample size bias effect is known and well understood for stationary OU models sampled without measurement error \cite{Tang2009a}. 
If both the time series data and innovation covariance are effectively stationary and mean zero, the  bias correction technique introduced in 
\cite{Calderon2013} for the KF (applicable to data observed with measurement error) can be heuristically applied to the MBF estimates \footnote{Note that the MBF has a time integrated measurement (this feature was not considered in Ref. \cite{Calderon2013}), hence the bias correction outlined in Ref. \cite{Calderon2013} may require modification in certain very high motion blur regimes (however, in the parameter regimes explored in this work, the bias correction for $\kappa$ was accurate.}.  After applying the correction outlined in Ref. \cite{Calderon2013}, the bias corrected parameters  of the MBF estimates are shown to coincide more closely with the true data generating process's $\kappa$.  
We stress that, in this example, we started in the stationary distribution (having mean zero) and the measurement noise did not vary over time. 
When the technical conditions hold for bias correction, the expectation of the parameter improves on average.  Also note that the bias correction is derived for the expected value obtained when averaging over multiple trajectories of length $T$.  
The bias correction improves performance on average \cite{Tang2009a,Calderon2013}, but it has a probability of degrading estimates even when all technical conditions required to apply the correction hold.   If the data is  confined (or is ``mean reverting'' \cite{Calderon2013}) around a nonzero mean at steady state, subtracting the empirical mean is a pragmatic way of ``enforcing'' the $v=0$ condition analyzed in Ref. \cite{Calderon2013}.  Inherently non-stationary finite trajectory length bias correction requires additional research 
\footnote{A general heuristic of potential pragmatic value in SPT should be noted regarding this statement: if particles are approaching their steady state distribution from ``non-equilibrium conditions'', bias in the estimated $\kappa$ tends to be small since there is more ``relaxation'' information encoded in the time series sequence and the practical utility of a bias correction scheme is reduced as can be verified empirically using the code provided.}.

\begin{figure}[H]
\centering
\begin{minipage}[b]{1.\linewidth}
\def\pw{1}
% \begin{overpic}[width=\pw\textwidth]{./Fig6.pdf}
\begin{overpic}[width=\pw\textwidth]{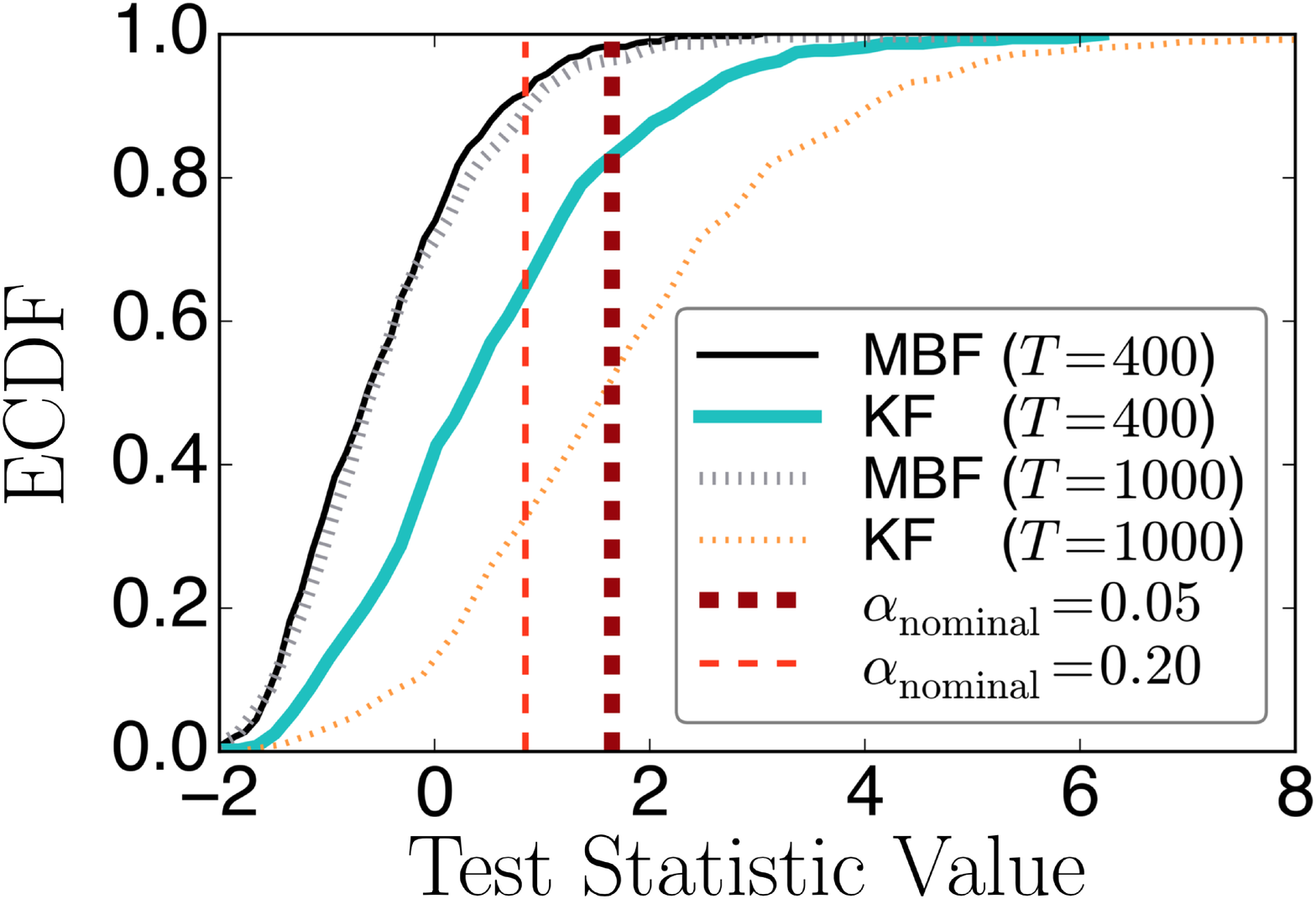}
\end{overpic}
\end{minipage}
\caption{\footnotesize \textbf{Statistically Testing Model Assumptions without Ground-truth.} Empirical cumulative distribution function (ECDF) obtained by evaluating the M(1,1) test statistic \cite{hong} at the MLE obtained $N$ using length $T$ simulations obtained with $\kappa=1 s^{-1}, \ D=0.1 \mu^2 /s , \   
 \sigma_{\mathrm{loc}} = 30 nm, \ \delta = 25 ms, \ v=0$.  Data was generated using motion blurred measurements.  The MLE and the corresponding M(1,1) statistic of the MBF (correctly modeling motion blur, hence the ``null model'') and the  KF (not modeling motion blur, hence representing ``model misspecificaiton'') were computed for each of the $N$ trajectories analyzed.  The length of the trajectories $T$ was increased to show the increase in power one can obtain in this regime.  Two  critical values corresponding to Type I error rates $\alpha$  of the large $T$ test statistic distribution are shown as vertical lines  (see text for additional details).
}
\label{fig:ECDF}
\end{figure}

In Fig. \ref{fig:ECDF}, we use the innovation sequence computed at the MLE to test the quality of the model via goodness-of-fit tests \cite{SPAgof,Calderon2013b,arxivDec2013}.  When analyzing experimental live cell data, one rarely has the luxury of ``ground truth'', so checking modeling assumptions against data is an important step.  
Here, we attempt to see if the correlation induced by the (simulated) motion blur can be detected when the classic KF is applied to blurred data. For this purpose, we re-analyzed the $\delta=50ms$ case shown in Fig. \ref{fig:diffProfile} and computed the $M(1,1)$ test statistic \cite{hong,Calderon2013b}.   The $M(1,1)$ test statistic aims to check if the conditional mean and correlation structure of the generalized residual series is consistent with that of a correctly specified model \cite{hong} (ignoring effects of motion blur primarily affects correlation in the generalized residual series).

 We plugged in the MLE for the KF and MBF (recall that the same trajectory was fit with multiple estimators) and used the data to compute the  $M(1,1)$ statistic for the $N=400$ trajectories of length $T=400$. The empirical cumulative distribution function (ECDF) of the $N=400$ test statistics is displayed for the two estimators. The vertical dashed lines plot the critical values corresponding to the limit normal null distribution of the $M(1,1)$ statistic \cite{hong}.  
The fraction of tests statistics  greater than these critical values can be rejected at a nominal Type I error rate  \footnote{``Nominal'' is used since the normal distribution requires $T \rightarrow \infty$ for these to be the true Type I error rates $\alpha$ \cite{hong}.} indicated by the graph. By inspecting  the intersection of the vertical lines with the ECDF, the fraction of the $N$ trajectories rejected for a nominal $\alpha$ can be  determined.   We simply picked a ``conservative'' and ``liberal'' rejection threshold to plot, however researchers can use the information encoded in the ECDF to carry out a test at any nominal Type I error level. 
For example, if one selected $\alpha_{\mathrm{nominal}} = 0.20$, $\approx 40\%$ of the KF fits are rejected when motion blurred data is fit with a model not accounting for the effects of motion blur.  As  $T$ increases, the statistical power (ability to reject  if the observed data is inconsistent with the assumed model) increases whereas the test statistics computed using the MBF innovation likelihood exhibit rejection rates just below the expected Type I error rates.  To illustrate how power increases with $T$, we show results obtained using the same parameters, but increasing trajectory length to $T=1000$.  
Using $\alpha_{\mathrm{nominal}} = 0.20$, $\approx 70\%$ of the KF fits are rejected with the increased $T$ value.

 Before concluding, we make some technical notes.  For  likelihood based time series analysis, it is recommended that a ``reasonable'' number of observations are used to estimate parameters 
 (accuracy depends on a variety of factors including $\delta$, $\theta$, $T$, etc.)  
 Some guidance about parameter accuracy and variability in the measurement noise free case can be obtained from probability and statistical theory \cite{Tang2009a}, but much theory is asymptotic in nature.
Using simulations in the parameter regime of interest to quantify the bias and MLE parameter variability is recommended. The interested reader can  tweak these parameters in the supplied IPython Notebooks to explore different regimes. At one extreme, 
if the product of $\kappa$ and $\delta$ is ``large" relative to the spatial and temporal resolution afforded by the measurement device, then detecting the temporal correlations in the time series data will be problematic with finite $T$ under the model assumed in Eq. \ref{eq:SDE}.  In this setting, using the parameterization used here, an MLE algorithm will typically estimate the stationary variance correctly,  
$\frac{\hat{D}}{\hat{\kappa}}$, but the individual components may not be reflective of the underlying truth.  
 At the other extreme, when the true $\kappa$ is near zero, other well-known technical problems occur due to so-called ``unit root'' technical complications arise \cite{hamilton}.  
In SPT terms, this effectively means no appreciable confinement can be detected and the particle may be exhibiting simple ``free" or ``directed'' diffusion.  Hence if the MBF analysis predicts $\kappa \le 0$ within statistical uncertainty, appropriate caution should be taken.

In live cell data, it is not expected that simple pure ``free'' or ``directed''
diffusion exist.  
Some  degree of confinement (due to the inherent crowded nature of the cell)  is almost always expected to be experimentally detectable in mobile particles tracked \emph{in vivo} with the resolution afforded by modern optical microscopes.   
Despite the technical caveats stated above, we have demonstrated that high accuracy parameter estimates can be obtained for fairly wide range of $\kappa$'s $\delta$'s and $D$'s relevant to SPT with reasonably ``small'' $T$ using the MBF.

\section{Conclusions}
\label{sec:conclusion}

The Motion Blur Filter (MBF) algorithm  was shown to be capable of consistently estimating  parameters required for extracting forces and diffusion coefficients given a single trajectory contaminated by static and dynamic measurement errors.  The approach can handle
  the  three most popular SPT models 
(confined, directed, and ``pure'' diffusion). 
The approach can consistently estimate molecular motion parameters from individual trajectories (enabling quantification of heterogeneity) in situations where the diffusion coefficients, $D$, span four orders of magnitude and the camera exposure times range from 5-100 $ms$ in the presence of confinement.    As discussed in Sec. \ref{sec:physicalSDE} and elsewhere \cite{Voisinne2010,Calderon2013,Hoze2014,CalderonBloom2015,Beheiry2015}, accurate and unbiased estimation of $D$ is important in  obtaining spatially 
 dependent effective molecular forces from position vs. time SPT data.

It was demonstrated that state-of-the-art estimators cannot consistently estimate motion parameters due either to neglected  motion blur or confinement effects. Using state-of-the-art estimators which do not explicitly model confinement and motion blur results in substantial biases of  $D$ (hence affecting estimates of molecular forces).
Other pragmatic issues  arising when analyzing individual trajectories, e.g. how to account for time varying localization accuracy, how to correct for  parameter  bias   encountered when trajectories are discretely sampled with finite length samples (finite sample size bias is prevalent in time series estimation \cite{Calderon2013}), and how to test fitted models against data with the MBF were discussed and demonstrated.
To facilitate implementation and promote reproducible research, we have provided Python scripts and IPython Notebooks on GitHub (this code both demonstrates general use and can also reproduce Figs. \ref{fig:nonstatVelocityEG} - \ref{fig:RvaryingEG}).

The approach, as presented, processed individual trajectories since a variety of microscopy techniques are now capable of producing long, high time resolution data from different imaging modalities \cite{Arhel2006,Manley2008,Thompson2010,Wells2010,Ram2012,Chen2014,Welsher2014,Backlund2014,Li2015,Wu2015,Moens2015}.  Both the localization quality and temporal resolution can deviate substantially from theory \cite{Thompson2002,Ober2004,Savin2005,Abraham2010}  (or even vary over time) and depend on the modality 
(e.g., the fluorescence channel in multicolor experiments can have different quality).  Our algorithm can  handle these practical complications faced by SPT researchers and produce output from different experiments which aims at removing experimental measurement and sampling artifacts to produce motion parameters representative of the true underlying tagged particle.    As we demonstrated in this work, a large source of bias introduced by the experimental apparatus is ``motion blur''. The software provided can be used to empirically explore different regimes of interest and the accuracy afforded by the MBF approach.  If a researcher desires to estimate one of the three  popular SPT motion models (confined/corralled, directed, ``pure'' diffusion, or some combination of these), it is recommended to use the MBF since the algorithm has demonstrated accuracy in many regimes of relevance to SPT (whereas other estimators are biased in some regimes).

If it is discovered  or believed that the parameters characterizing the dynamics driving the motion of the molecule(s) of interest at a given spatial location in a cell are independent of time \cite{Manley2008,Hoze2014,Beheiry2015}, one can  modify our algorithm to aggregate multiple time series even if they have vastly different localization precisions or exposure times. For example, one could use each trajectory to produce a likelihood function (given trajectory specific localization information) and then develop a cost function which aims to find the single parameter vector minimizing the net sum of the log likelihoods.  However, this trajectory aggregation requires a strong assumption regarding spatial and temporal stationarity and we believe that potentially interesting transient molecular events will be missed by this type of approach \cite{Calderon2013b,CalderonBloom2015}, hence we advise researchers to start by analyzing data on a trajectory-wise basis \cite{SPAfilter,SPAdsDNA}. In addition to parameter estimation, the likelihood based scheme provides diagnostic statistics which can be used to check statistical modeling assumptions directly against data via goodness-of-fit tests without ``ground truth'' (checking both shape and/or statistical dependence assumptions implicit in the model \cite{SPAgof,Calderon2013b,arxivDec2013}).  We demonstrated the ability of these tests to detect unmodeled correlations in the classic KF induced by motion blur effects.  However, the same hypothesis testing procedure can also be used to determine whether the assumptions required to carry out the ``trajectory aggregation'' mentioned above (e.g., use many different trajectories to estimate a parameter vector characterizing the dynamics at a fixed spatial location) are justified by the empirical data.

The MBF estimator leveraged signal processing and stochastic process ideas to synthesize a new algorithm capable of addressing many open practical issues facing SPT data analysis.  
The dynamical model underlying  the MBF is a continuous time linear  SDEs driven by ``standard Brownian diffusion'' \cite{kp}.
As stated in the Introduction, ``anomalous diffusion" can result when one averages over many types of dynamical states \cite{Hofling2013,Metzler2014}, however resolution afforded by contemporary microscopes permits temporal resolution where  standard diffusion models are useful.  A primary aim of this work is to provide a computational tool which can be leveraged when the molecular events of interest occur within the spatial and temporal resolution of optical microscopes before signatures of ``anomalous diffusion" manifest themselves in the data.  Under these conditions, backing out 
``effective forces" from the local molecular diffusivity is reasonable \cite{Calderon2013b,CalderonBloom2015,Holcman2015}.  
For longer trajectories, this may require one to segment trajectories into distinct kinetic states 
\cite{CalderonBloom2015} and then apply the analysis to the segments.  We did not present segmentation results, but the likelihood based MBF algorithm can be used to modify the cost function of existing state-of-the-art time series segmentation algorithms \cite{Fox2011,CalderonBloom2015} and remove artifacts induced by motion blur and unknown  localization noise (these noise sources are ubiquitous in SPT data analysis). 

Explicitly accounting for motion blur is also expected to facilitate segmenting data where multiple imaging modalities (where data is acquired with different temporal resolutions and/or exposure times) are combined to describe the dynamics of individual molecules in living cells.   We presented  1D (scalar) illustrative examples, but the MBF algorithm can  process multivariate signals.  
However, obtaining closed-form expressions for the filter quantities is slightly complicated by matrix exponentials. The computational challenges with multivariate extensions is left to future work. 
 
% \ \\
%  \noindent \textbf{\large Acknowledgments:} 
\begin{acknowledgments}
 The author thanks Kerry Bloom for helpful motivating discussions and comments related to this as well as Kerry Bloom, Scott Lundberg, and two anonymous referees who provided valuable comments which improved the exposition of this paper.  This research was supported by internal R\&D funds of Ursa Analytics, Inc.
\end{acknowledgments}

% {\footnotesize
	
\section{Appendix}

The well-known solution to the Ornstein-Uhlenbeck (OU)  SDE (Eq. \ref{eq:SDE}) can be written explicitly as:
\begin{align}
r_\delta  = & A + Fr_0 + \int_0^\delta \exp\big(\kappa(s-\delta)\big)\sqrt{2D} dB_s,
\label{eq:FA}
\end{align}

\noindent where $A \equiv \big(1-\exp(-\delta \kappa)\big)\frac{v}{\kappa} $ and $F \equiv \exp(-\delta \kappa)$ \cite{kp}.
The solution above can be written in terms of simple Gaussian random variables since the OU process is one of the rare cases where a SDE can be solved in closed-form by appealing to integration factor techniques  used in standard ordinary differential equations. ``Solved" means the process can be written explicitly in terms of time and a Brownian motion path. When an SDE is ``solved", realizations can be constructed without numerical integration approximations of any deterministic or stochastic integrals \cite{kp}.  
The expression above can be used to compute closed-form expressions for: 
\begin{align}
\mathbb{E}[r_\delta | r_0] = & A + Fr_0 \\ 
\label{eq:HFA}
\mathbb{E}[\frac{1}{\delta}\int_0^\delta r_\delta | r_0] 
= & H_A + H_Fr_0  \\
H_F = & \frac{1}{\delta}\int_0^\delta \exp(-s \kappa)ds \\
\nonumber &  =  \frac{1}{\kappa\delta}(1 - \exp(-\delta\kappa) ) \\
H_A = & \frac{1}{\delta}\int_0^\delta \big(1-\exp(-s \kappa)\big)\mu ds  \\
\nonumber & =  \mu - \frac{\mu}{\delta\kappa} + \frac{\mu \exp(-\delta\kappa)}{\delta\kappa} \\
\mu := & \frac{v}{\kappa} \\
\end{align}

\noindent where  $\delta>0$ (note all discrete parameters defined in this Appendix depend implicitly on $\delta$). 

% `"The expressions just derived provide the means of the state and measurement forecast in Algorithm 1.
Since the OU process considered is linear and driven by standard Brownian motion, it is characterized by the first two moments and covariances of the process.  The means have been defined above. The second moments and covariances  can also be computed as an explicit  function of time and $\theta$ for the model considered. Recall that 
the KF and MBF uncertainty estimates both are centered around using covariances of mean zero estimates to make various linear projections \cite{anderson1979}.  In what follows, without loss of generality, we assume $r_0$ is statistically independent of 
$B_t \ \forall t > 0$ and that $r_0 = 0$, $v=0$ (so the relation   
$\mathrm{cov}(r_t, r_s) =  \mathbb{E}[r_tr_s]  $ holds for $s\le t$), and $\kappa \ge 0$. Under these non-restrictive conditions and  using the well-known quadratic variation properties of Brownian motion \cite{kp}, 
one has the following closed-form relationship for the state covariance \cite{kp}: 
% \begin{align}
% % \mathrm{cov}(r_t, r_s) : = &
% % \mathbb{E}[\int_0^t \exp\big(\kappa(u-t)\big)\sqrt{2D} dB_u\int_0^s \exp\big(\kappa(v-s)\big)\sqrt{2D} dB_v] \\
%  % = & 2D\int_0^s \exp\big(\kappa(v-t)\big)\exp\big(\kappa(v-s)\big)dv \\ 
%  % = & \frac{D}{\kappa} \big(\exp(2\kappa s) - 1 \big)\exp(-\kappa (s+t)).
%  \label{eq:covQ}
% \end{align}
\ \\ %unsure of how PRE wants to typeset this....make look nice 
\ \\
\ \\
\begin{flalign}
 \nonumber \mathrm{cov}(r_t, r_s) & :=   & \\
 \label{eq:covQ}
\nonumber & \mathbb{E}[\int_0^t \exp\big(\kappa(u-t)\big)\sqrt{2D} dB_u\int_0^s \exp\big(\kappa(v-s)\big)\sqrt{2D} dB_v] &  \\
\nonumber & =  2D\int_0^s \exp\big(\kappa(v-t)\big)\exp\big(\kappa(v-s)\big)dv & \\ 
 & =  \frac{D}{\kappa} \big(\exp(2\kappa s) - 1 \big)\exp(-\kappa (s+t)). & 
\end{flalign}
The above expression is valid for $s \le t$.
The state  covariance  above can be used to solve for the variance and covariance of other quantities required by the MBF.  Specifically one needs to compute both $\mathbb{E}[\frac{r_\delta }{\delta} \int^\delta_0 r_tdt]$ and 
$\mathbb{E}[\frac{1}{\delta}\int^\delta_0 r_tdt \times \frac{1}{\delta}\int^\delta_0 r_sds]$ 
to solve the ``Corrector'' update (see Fig. \ref{fig:flowChart}).  The former expectation provides $\mathrm{cov}(r_{i},\tilde{\psi}_i)$ and the latter 
provides the contribution of motion blur, $\mathrm{cov}(\epsilon^{\mathrm{mblur}}_{t_i},\epsilon^{\mathrm{mblur}}_{t_i})$, to the net measurement covariance $\mathrm{cov}(\tilde{\psi}_i,\tilde{\psi}_i)$ stated in Eq. \ref{eq:KFfilterUpdate}.
Note that in the MBF where uniform illumination is assumed, we 
use both $\delta$ and $t_E$ to represent the exposure time.  The order of time integration and expectation can be exchanged for $\delta > 0 $ for the process considered, e.g.  
$\mathbb{E}[\frac{r_\delta }{\delta} \int^\delta_0 r_tdt] = \frac{1}{\delta} \int^\delta_0 \mathbb{E}[r_\delta r_t]dt$ \cite{kp}.  This reduces the problem to solving standard time integrals, since $\mathbb{E}[r_tr_s]=\mathrm{cov}(r_t, r_s)$ (recall our non-restrictive 
assumptions on the process mean) has already been solved in terms of $\theta,t$ and $s$ as shown above.  
With this background, it can also be shown that: 

\begin{align}
\delta = & t_{i} - t_{i-1} = t_E \ \forall i \\
Q^{\mathrm{mblur}} := & \mathrm{cov}(\epsilon^{\mathrm{mblur}}_{t_i},\epsilon^{\mathrm{mblur}}_{t_i}) \\
= & \frac{D}{\kappa\delta^2}\big(  
\frac{2 \delta}{\kappa} - \frac{3}{\kappa^{2}} + \frac{4}{\kappa^{2} e^{\delta \kappa}} - \frac{1}{\kappa^{2} e^{2 \delta \kappa}}\big) \\
C = \ & \mathrm{cov}(\eta_{t_{i-1}},\epsilon^{\mathrm{mblur}}_{t_{i}}) \\
    = & \frac{D}{\kappa \delta}\big(\frac{1}{\kappa} - \frac{2\exp(-\kappa \delta)}{\kappa} + \frac{\exp(-2\kappa \delta)}{\kappa}\big)
\end{align}
The first line in the set of equations above is to remind the reader that  it is assumed no ``missing frames" exist in the uniform illumination measurement model stated in Eq. \ref{eq:contMeas}.  Missing frames may occur in time lapse experiments or if quantum dots blink  (both missing frames and more exotic shutter functions could be considered within the MBF framework, but these cases require a more complicated notation which we have elected not to explore in this article which introduces the basic MBF). Combining the expressions for $(Q^{\mathrm{mblur}},C)$ above, with the expressions for $(A,F)$ reported previously (see below Eq. \ref{eq:FA}), $Q=\mathrm{cov}(r_\delta, r_\delta)$ (see Eq. \ref{eq:covQ}) and the expressions for $(H_F,H_A)$  (see Eq. \ref{eq:HFA}) provide the closed-form expressions required to precisely discretize the model reported in Eqs. \ref{eq:SDE} and \ref{eq:contMeas} without any statistical approximations in the filtering framework.  We remind the reader that all the quantities derived depend implicitly on both $\delta$ and 
%the parameters in 
$\theta$.

% \bibliography{running,biomedtracking,mRNAII,StatPapers,mRNA,Primary_Cilium,tracking,HIV_Gag,biomedtracking-3C,chemical_physics,Primary_Cilium_Shared,4DLiveCell_ChromosomeDynamics} 

% merlin.mbs apsrev4-1.bst 2010-07-25 4.21a (PWD, AO, DPC) hacked
% Control: key (0)
% Control: author (8) initials jnrlst
% Control: editor formatted (1) identically to author
% Control: production of article title (-1) disabled
% Control: page (0) single
% Control: year (1) truncated
% Control: production of eprint (0) enabled
%

\end{document}